\documentclass[pdftex,twocolumn,epjc3]{svjour3}          

\usepackage[none]{hyphenat}
\smartqed  
\usepackage{mathtools, cuted, widetext}
\usepackage{amsmath,epsfig,amssymb,multirow}
\RequirePackage{graphicx}
\RequirePackage{mathptmx}      
\RequirePackage{flushend}
\RequirePackage[numbers,sort&compress]{natbib}
\RequirePackage[colorlinks,citecolor=blue,urlcolor=blue,linkcolor=blue]{hyperref}

\def\p{\partial}
\def\wtil#1{\widetilde{#1}}
\begin{document}
\sloppy
	\title{On polarization parameters of spin-$1$ particles and anomalous 
		couplings in {\boldmath $e^+e^-\to ZZ/Z\gamma$}}
	\author{Rafiqul Rahaman\thanksref{e1,adrss} \and Ritesh K. Singh\thanksref{e2,adrss}}
	\thankstext{e1}{email:rr13rs033@iiserkol.ac.in}
	\thankstext{e2}{email:ritesh.singh@iiserkol.ac.in}
	\institute{Department of Physical Sciences,
		Indian Institute of Science Education and Research Kolkata,
		Mohanpur, 741246, India\label{adrss}}
\date{}
\maketitle

\begin{abstract}
We study the anomalous trilinear gauge couplings of $Z$ and $\gamma$ using a
complete set of polarization asymmetries for the $Z$ boson in 
$e^+e^-\to ZZ/Z\gamma$ processes with unpolarized initial beams.
We use these
polarization asymmetries, along with the cross section, to obtain a simultaneous
limit on all the anomalous couplings using the Markov-Chain--Monte-Carlo (MCMC) 
method. For an $e^+e^-$ collider running at $500$ GeV center-of-mass energy and
$100$ fb$^{-1}$ of integrated luminosity the simultaneous limits on the 
anomalous couplings are $1\sim3\times 10^{-3}$.
\end{abstract}

\section{Introduction}
The Standard Model (SM) of particle physics is a well-established theory and 
its particle spectrum has been completed with the discovery of the Higgs 
boson~\cite{Chatrchyan:2012xdj} at the Large Hadron Collider (LHC). 
The predictions of the SM are being confirmed  time and again in various 
colliders with great success, and yet phenomena such as CP-violation, 
neutrino oscillation, baryogenesis, dark matter,  etc., require one to look 
beyond the SM. Most of the beyond SM (BSM) models need either new particles or
new couplings among the SM particles or both. Often this leads to a modified
electro-weak sector with modified couplings. To test the SM (or BSM) predictions
for the electro-weak symmetry breaking (EWSB) mechanism one needs precise
measurements of the strength and tensorial structure of the Higgs ($H$) 
coupling 
with all other gauge boson ($W^\pm$, $\gamma$, $Z$), Higgs self-couplings and 
couplings among the gauge boson themselves.

In this work, we focus on the precise measurement of trilinear gauge boson 
couplings, in a model independent way, at the proposed International Linear 
Collider (ILC)~\cite{Behnke:2013xla,Baer:2013cma}.
The possible trilinear gauge boson interactions in electro-weak (EW) theory 
are $WWZ$, $WW\gamma$, $ZZ\gamma$, $ZZZ$, $\gamma\gamma Z$, and 
$\gamma\gamma\gamma$, out of which the SM, after EWSB,  provides  only $WWZ$ 
and  $WW\gamma$ self-couplings. Other interactions among neutral gauge bosons
are not possible at the tree level in the SM and hence they are anomalous. 
Thus any deviation from the SM prediction, either in strength or the tensorial
structure, would be a signal of  BSM physics. There are two ways to study
these anomalous couplings in a model independent way: The first way is to write down
an effective vertex using the most general set of tensorial structures for it 
satisfying Lorentz invariance, $U(1)_{em}$ invariance, and Bose symmetry weighted 
by corresponding CP-even/odd form factors~\cite{Gaemers:1978hg,
Renard:1981es, Hagiwara:1986vm}. There is, a priori, no relation between various
form factors. The other way to study anomalous couplings is to add a set of 
higher-dimension operators, invariant under (say)
$SU(2)_L\otimes U(1)_Y$~\cite{Buchmuller:1985jz}, to the SM Lagrangian and then
obtain the effective vertices with anomalous couplings after EWSB.
This method has been used to study anomalous triple neutral 
gauge boson couplings with the SM gauge 
group~\cite{Larios:2000ni,Cata:2013sva,Degrande:2013kka}.

The neutral sector of the anomalous vertices discussed in~\cite{Hagiwara:1986vm}
include $ZZZ$, $ZZ\gamma$, and $Z\gamma\gamma$ interactions and 
have been widely studied in the literature 
\cite{Czyz:1988yt,Boudjema:Desy1992,Baur:1992cd,Choudhury:1994nt,Choi:1994nv,
Aihara:1995iq,Ellison:1998uy, Gounaris:1999kf,Gounaris:2000dn,Baur:2000ae,
Rizzo:1999xj,Atag:2003wm,Ananthanarayan:2004eb,Ananthanarayan:2011fr,
Ananthanarayan:2014sea,Poulose:1998sd} in the context of different collider; 
$e^+e^-$ collider~\cite{Czyz:1988yt,Boudjema:Desy1992,Choudhury:1994nt,
Ananthanarayan:2004eb,Ananthanarayan:2011fr,Ananthanarayan:2014sea}, hadron 
collider~\cite{Baur:1992cd,Ellison:1998uy,Baur:2000ae}, both $e^+e^-$ and 
hadron collider~\cite{Aihara:1995iq,Gounaris:1999kf,Gounaris:2000dn}, $e\gamma$ 
collider~\cite{Choi:1994nv,Rizzo:1999xj,Atag:2003wm} and $\gamma\gamma$ 
collider~\cite{Poulose:1998sd}.
For these effective anomalous vertices one can write an effective Lagrangian
and they have been given in \cite{Boudjema:Desy1992, Gounaris:1999kf,
Choi:1994nv, Choudhury:1994nt} up to differences in conventions and
parametrizations. The Lagrangian corresponding to the anomalous form factors in 
the neutral sector in \cite{Hagiwara:1986vm} is given 
by~\cite{Gounaris:1999kf}:
\newpage
\begin{widetext}
\begin{eqnarray}
{\cal L}_{aTGC} &&= \frac{g_e}{M_Z^2} \Bigg [
-[f_4^\gamma (\partial_\mu F^{\mu \beta})+
f_4^Z (\partial_\mu Z^{\mu \beta}) ] Z_\alpha
( \partial^\alpha Z_\beta)+
[f_5^\gamma (\partial^\sigma F_{\sigma \mu})+
f_5^Z (\partial^\sigma Z_{\sigma \mu}) ] \wtil{Z}^{\mu \beta} Z_\beta
\nonumber \\
&-&  [h_1^\gamma (\partial^\sigma F_{\sigma \mu})
+h_1^Z (\partial^\sigma Z_{\sigma \mu})] Z_\beta F^{\mu \beta}
-[h_3^\gamma  (\partial_\sigma F^{\sigma \rho})
+ h_3^Z  (\partial_\sigma Z^{\sigma \rho})] Z^\alpha
 \wtil{F}_{\rho \alpha}
\nonumber \\
&- & \left \{\frac{h_2^\gamma}{M_Z^2} [\partial_\alpha \partial_\beta
\partial^\rho F_{\rho \mu} ]
+\frac{h_2^Z}{M_Z^2} [\partial_\alpha \partial_\beta
(\square +M_Z^2) Z_\mu] \right \} Z^\alpha F^{\mu \beta}
+ \left \{
\frac{h_4^\gamma}{2M_Z^2}[\square \partial^\sigma
F^{\rho \alpha}] +
\frac{h_4^Z}{2 M_Z^2} [(\square +M_Z^2) \partial^\sigma
Z^{\rho \alpha}] \right \} Z_\sigma \wtil{F}_{\rho \alpha }
 \Bigg ] ~ ,\nonumber\\ \label{LaTGCfull}
\end{eqnarray}
\end{widetext}
where $\wtil{Z}_{\mu \nu}=1/2 \epsilon_{\mu \nu \rho \sigma}Z^{\rho
\sigma}$ ($\epsilon^{0123}=+1$)  with
$Z_{\mu\nu}=\partial_\mu Z_\nu -\partial_\nu Z_\mu$ and similarly for
the photon tensor $F_{\mu\nu}$. The couplings $f_4^V, ~h_1^V,~ h_2^V$
correspond to the CP-odd tensorial structures, while $f_5^V, ~h_3^V ,~ h_4^V$
correspond to the CP-even ones. Further, the terms 
corresponding to $h_2^V$ and $h_4^V$ are of mass dimension-$8$,  while the others are dimension-$6$ in the Lagrangian.
In \cite{Ananthanarayan:2014sea} the authors have pointed out one more possible 
dimension-$8$ CP-even term  for $Z \gamma Z $ vertex, given by
$${\cal L}_{aTGC} \supset \dfrac{g_e h_5^Z}{2M_Z^4} (\p^\tau F^{\alpha
\lambda})\wtil{Z}_{\alpha\beta}\p_\tau\p_\lambda Z^\beta.$$ 
In our present work, however,  we shall restrict ourselves to the
dimension-$6$ subset of the Lagrangian given in Eq.~(\ref{LaTGCfull}).

On the theoretical side, 
it is possible to generate some of these anomalous tensorial structures within
the framework of a renormalizable theory at higher loop orders, for example,
through a fermion triangle diagram in the SM. Loop generated anomalous couplings
have been studied for the Minimal Supersymmetric SM 
(MSSM)~\cite{Gounaris:2000tb,Choudhury:2000bw} and Little Higgs 
model~\cite{Dutta:2009nf}. Beside this, a non-commutative extension of the SM
(NCSM)~\cite{Deshpande:2001mu,Deshpande:2011uk} can also provide a anomalous
coupling structure in the neutral sector with a possibility of a trilinear 
$\gamma \gamma \gamma$ coupling as well~\cite{Deshpande:2001mu}.

On the experimental side, the anomalous Lagrangian in Eq.~(\ref{LaTGCfull})
has been explored at the LEP~\cite{Acciarri:2000yu,Abbiendi:2000cu,
Abbiendi:2003va,Achard:2004ds,Abdallah:2007ae}, the 
Tevatron~\cite{Abazov:2007ad,Aaltonen:2011zc,Abazov:2011qp}, and the
LHC~\cite{Chatrchyan:2012sga,Chatrchyan:2013nda,Aad:2013izg,
Khachatryan:2015kea,Khachatryan:2016yro}. The tightest bounds  on 
$f_i^V$($i=4,5$)~\cite{Chatrchyan:2012sga} and on 
$h_j^V$($j=3,4$)~\cite{Khachatryan:2016yro} comes
from the CMS collaboration (see Table~\ref{tab:aTGC_constrain_form_collider}). 
For the $ZZ$ process the total rate has been 
used~\cite{Chatrchyan:2012sga}, while for the  $Z\gamma$ process both
the cross section and the $p_T$ distribution of $\gamma$ has been 
used~\cite{Chatrchyan:2013nda,Aad:2013izg,Khachatryan:2015kea,
Khachatryan:2016yro} for obtaining the limits. All these analyses vary one parameter at
a time to find the $95$~\% confidence limits on the form factors. For the $Z\gamma$ process
the limits on the CP-odd form factors, $h_1^V, \ h_2^V$, are comparable to the 
limits on the CP-even form factors, $h_3^V, \ h_4^V$, respectively.
\begin{table}
\centering
\caption{\label{tab:aTGC_constrain_form_collider} List of tightest limits on 
anomalous couplings of Eq.~(\ref{LaTGCfull}) available in literature}
\renewcommand{\arraystretch}{1.5}
\begin{tabular*}{\columnwidth}{@{\extracolsep{\fill}}lll@{}}\hline
Coupling & Limits & Remark\\ \hline
$f_4^\gamma$ & $0.0_{-1.3\times 10^{-2}}^{+1.5\times 10^{-2}}$ & LHC\\
$f_4^Z$ &  $0.0_{-1.1\times 10^{-2}}^{+1.2\times 10^{-2}}$ & $7$ TeV\\
$f_5^\gamma$ & $0.0\pm 1.4 \times 10^{-2}$&  $5$ fb$^{-1}$\\
$f_5^Z$ & $0.0\pm  1.2 \times 10^{-2}$ & Ref.~\cite{Chatrchyan:2012sga}\\ \hline
$h_3^\gamma$  & $0.0_{-1.1\times 10^{-3}}^{+0.9\times 10^{-3}}$ & LHC \\
$h_3^Z$ & $0.0_{-1.5\times 10^{-3}}^{+1.6\times 10^{-3}}$ & 8 TeV\\
$h_4^\gamma$  & $0.0_{-3.8 \times 10^{-6} }^{+ 4.3 \times 10^{-6}} $ & $19.6$
fb$^{-1}$\\
$h_4^Z$ & $0.0_{-3.9 \times 10^{-6} }^{ +4.5 \times 10^{-6} } $ &
Ref.~\cite{Khachatryan:2016yro}\\
\hline
\end{tabular*}
\end{table}

To put simultaneous limits on all the form factors one would need as many 
observables as possible, like differential rates, kinematic asymmetries, etc. 
A set of asymmetries with respect to the initial
beam polarizations has been considered~\cite{Czyz:1988yt,Choudhury:1994nt,
Choi:1994nv,Gounaris:1999kf,Rizzo:1999xj,Atag:2003wm,Ananthanarayan:2004eb,
Ananthanarayan:2011fr,Ananthanarayan:2014sea} at the $e^+e^-$ and $e\gamma$
colliders.  References~\cite{Czyz:1988yt,Choi:1994nv,Ananthanarayan:2011fr,
Ananthanarayan:2014sea} also include CP-odd asymmetries made out of the initial
beam polarizations alone, which are instrumental in putting strong limits on
CP-odd form factors. Additionally, Ref.~\cite{Czyz:1988yt} includes the
polarization of produced $Z$ as well for forming the asymmetries. The latter does
not necessarily require initial beams to be polarized and can be generalized to
the hadron colliders as well where such a beam polarization may not be available.

To this end, we discuss angular asymmetries in colliders 
corresponding to different polarizations of $Z$ boson in particular and any 
spin-$1$ particle in general.
For a spin-$s$ particle, the polarization density matrix is a 
$(2s+1)\times(2s+1)$ hermitian, unit-trace matrix that can be parametrized by
$4s(s+1)$ real parameters. These parameters are different kinds of polarization.
For example, a spin-$1/2$ fermion has three polarization parameters called
longitudinal, transverse, and normal polarizations (see for example
\cite{Godbole:2006tq,Boudjema:2009fz}).
Similarly, for a spin-$1$ particle we have a total of eight such parameters. Three
of them are vectorial like in the spin-$1/2$ case and the other five are 
tensorial~\cite{Boudjema:2009fz,Aguilar-Saavedra:2015yza} as will be described in
Section~\ref{sec:pol_obs} for completeness. Eight polarization
parameters for a massive spin-$1$ particle have been discussed earlier in the
context of anomalous trilinear gauge couplings~\cite{Ots:2004hk,Ots:2006dv}, for
the spin measurements studies~\cite{Boudjema:2009fz} and to study processes
involving $W^\pm$ bosons~\cite{Aguilar-Saavedra:2015yza}. In this work
we will investigate all anomalous couplings (up to dimension-$6$ operators) 
of Eq.~(\ref{LaTGCfull}) in the processes  $e^+e^-\to ZZ/Z\gamma$ with
the help of the total cross section and the eight polarization asymmetries of the final state $Z$ boson. 

The plan of this paper is as follows. In Section~\ref{sec:pol_obs} we
discuss the polarization observables of a spin-$1$ boson in detail using the
language of polarization density matrices. Section~\ref{sec:Lag} has a brief 
discussion of the anomalous Lagrangian and corresponding off-shell vertices
and the required on-shell vertices. In Section~\ref{sec:asym_sen} we study the
sensitivity of the polarization asymmetries to the anomalous couplings and
in Section~\ref{sec:MCMC} we perform a likelihood mapping of the full coupling
space using Markov Chain Monte Carlo (MCMC) method for two benchmark points
 along with a likelihood
ratio based hypothesis testing to resolve the two benchmark points. We conclude in
Section~\ref{sec:conclusion}.

\section{Polarization observables for spin-$1$ boson}
\label{sec:pol_obs}
In this section, we discuss the complete set of polarization observables of 
a spin-$1$ particle in a generic production process and the method 
to extract them from the distribution of its decay products. 
\begin{figure}[ht]
\centering
\includegraphics[width=7.0cm]{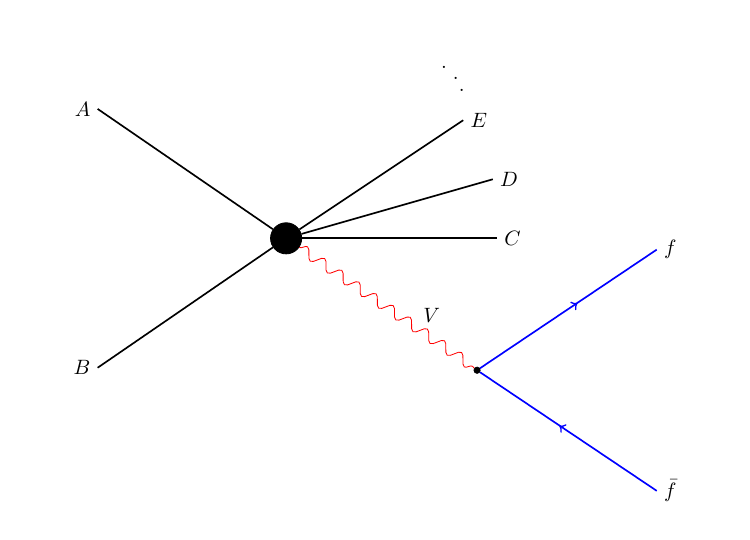}
\caption{\label{fig:Z_production_decay} Feynman diagram for a general process 
for production of spin-$1$ particle ($Z$ boson) and its decay to a pair of 
fermions} 
\end{figure}

\subsection{The production and decay density matrices}
We consider a generic reaction $A \ B \to V \ C \ D \ E \ldots$, where particle
$V$ is massive and has spin $1$, see Fig.~\ref{fig:Z_production_decay}.  
The production density matrix for particle $V$ can be written as
\begin{equation}\label{production_density_matrix}
\rho(\lambda,\lambda^\prime) = \frac{\mbox{Phase \ space}}{\mbox{Flux}} 
 \ \mathcal{M}(\lambda)  
\mathcal{M}^\dagger(\lambda^\prime).
\end{equation} 
Here, $\mathcal{M}(\lambda)$ is the helicity amplitude for the production of 
$V$ with  helicity  $\lambda\in\{-1,0,1\}$, while the helicities of all the 
other particles ($A, B, C, D, E\ldots$ ) are suppressed. 
The differential cross section for the production of $V$ would be given by
\begin{eqnarray}
\dfrac{d\sigma_V}{d\Omega_V}&=&\text{Trace}\left[\rho(\lambda ,\lambda^\prime)\right]
= \rho(+,+)+\rho(0,0)+\rho(-,-).\nonumber\\
\end{eqnarray} 
Here $\Omega_V$ is the solid angle of the particle $V$, while the phase space
corresponding to all other final state particles is integrated out.
The production density matrix can be written in terms of 
polarization density matrix $P(\lambda,\lambda^\prime)$ 
as~\cite{Boudjema:2009fz}
\begin{eqnarray}\label{Polarization_and_production_cross_section}
 P(\lambda ,\lambda^\prime)&=&\dfrac{1}{\sigma_V}\int\rho(\lambda ,\lambda^\prime)d\Omega_V
=\dfrac{1}{\sigma_V}{\rho}_T(\lambda ,\lambda^\prime).
\end{eqnarray}
Here the matrix $P(\lambda,\lambda^\prime)$ is a $3\times3$ hermitian matrix 
that can be parametrized in terms of a vector $\vec{p}=(p_x,p_y,p_z)$ and a 
symmetric, traceless, second-rank tensor $T_{ij}$ as
\begin{eqnarray}
\label{Polarization_matrix}
&&P(\lambda,\lambda') =\nonumber\\
\renewcommand{\arraystretch}{1.5}
&& \left[
\begin{tabular}{lll}
$\frac{1}{3}+\frac{p_z}{2}+\frac{T_{zz}}{\sqrt{6}}$ &
$\frac{p_x -ip_y}{2\sqrt{2}}+\frac{T_{xz}-iT_{yz}}{\sqrt{3}}$ &
$\frac{T_{xx}-T_{yy}-2iT_{xy}}{\sqrt{6}}$ \\
$\frac{p_x +ip_y}{2\sqrt{2}}+\frac{T_{xz}+iT_{yz}}{\sqrt{3}}$ &
$\frac{1}{3}-\frac{2 T_{zz}}{\sqrt{6}}$ &
$\frac{p_x -ip_y}{2\sqrt{2}}-\frac{T_{xz}-iT_{yz}}{\sqrt{3}}$ \\
$\frac{T_{xx}-T_{yy}+2iT_{xy}}{\sqrt{6}}$ &
$\frac{p_x +ip_y}{2\sqrt{2}}-\frac{T_{xz}+iT_{yz}}{\sqrt{3}}$ &
$\frac{1}{3}-\frac{p_z}{2}+\frac{T_{zz}}{\sqrt{6}}$
\end{tabular}\right].\nonumber\\
\end{eqnarray}
For a spin-$1$ particle decaying to a pair of fermions via the interaction
vertex $Vf\bar{f}: \gamma^\mu \left(L_f \ P_L + R_f \ P_R \right)$, the 
decay density matrix (normalized to one) is given by~\cite{Boudjema:2009fz}
\vspace{1.5cm}
\begin{widetext}
\centering
\begin{equation}
\renewcommand{\arraystretch}{1.5}
\Gamma(\lambda ,\lambda^\prime)=\left[
\begin{tabular}{lll}
$\frac{1+\delta+(1-3\delta)\cos^2\theta+2\alpha \cos\theta}{4}$ &
$\frac{\sin\theta(\alpha+(1-3\delta)\cos\theta)}{2\sqrt{2}} \ e^{i\phi}$&
$(1-3\delta)\frac{(1-\cos^2\theta)}{4} \ e^{i2\phi}$\\
$\frac{\sin\theta(\alpha+(1-3\delta)\cos\theta)}{2\sqrt{2}} \ e^{-i\phi}$&
$\delta+(1-3\delta)\frac{\sin^2\theta}{2}$ &
$\frac{\sin\theta(\alpha-(1-3\delta)\cos\theta)}{2\sqrt{2}} \ e^{i\phi}$\\
$(1-3\delta)\frac{(1-\cos^2\theta)}{4} \ e^{-i2\phi}$ &
$\frac{\sin\theta(\alpha-(1-3\delta)\cos\theta)}{2\sqrt{2}} \ e^{-i\phi}$ &
$\frac{1+\delta+(1-3\delta)\cos^2\theta-2\alpha\cos\theta}{4}$
\end{tabular} \right]. 
\label{decay_density_matrix1}
\end{equation}
\end{widetext}
Here $\theta$, $\phi$ are the polar and the azimuthal orientation of the  final 
state fermion $f$, in the rest frame of $V$ with its would be momentum along  
$z$-direction. The parameters $\alpha$ and $\delta$ are given by
\begin{equation}
\alpha=
\frac{2(R_f^2-L_f^2)\sqrt{1+(x_1^2-x_2^2)^2-2(x_1^2+x_2^2)}}
{12 L_fR_f x_1x_2+(R_f^2+L_f^2)[2-(x_1^2-x_2^2)^2+(x_1^2+x_2^2)]},
\end{equation}
\begin{equation}
\delta=\frac{4L_fR_f x_1x_2
+(R_f^2+L_f^2)[(x_1^2+x_2^2)-(x_1^2-x_2^2)^2]} {12 L_fR_f x_1x_2+(R_f^2+L_f^2)[2-(x_1^2-x_2^2)^2+(x_1^2+x_2^2)]},
\end{equation}
with $x_i=M_f/M_V$. For massless final state fermions,
$x_1\to0, \ x_2\to 0$, one obtains $\delta \to 0$ and 
$\alpha \to (R_f^2- L_f^2)/ (R_f^2+L_f^2)$. 
Combining the production and decay density matrices, the total rate for the
process shown in Fig.~\ref{fig:Z_production_decay}, with particle $V$ being
on-shell, is given by~\cite{Boudjema:2009fz},
\begin{equation}
\dfrac{1}{\sigma}\dfrac{d\sigma}{d\Omega_f}=\frac{2s+1}{4\pi}
\sum_{\lambda,\lambda^\prime}^{}P(\lambda,\lambda^\prime)
\Gamma(\lambda,\lambda^\prime),
\label{eq:norm_dist}
\end{equation}
where $\sigma=\sigma_V BR(V\to f\bar{f})$ is the total cross section for 
production of $V$ including its decay and $s=1$ being spin of the particle.
Using Eqs.~(\ref{Polarization_matrix}) and (\ref{decay_density_matrix1}) in
Eq.~(\ref{eq:norm_dist}), the  angular  distribution for the fermion $f$ 
becomes
\begin{eqnarray}
\frac{1}{\sigma} \ \frac{d\sigma}{d\Omega_f} &=&\frac{3}{8\pi} \left[
\left(\frac{2}{3}-(1-3\delta) \ \frac{T_{zz}}{\sqrt{6}}\right) + \alpha \ p_z
\cos\theta \right.\nonumber\\
&+& \sqrt{\frac{3}{2}}(1-3\delta) \ T_{zz} \cos^2\theta
 \nonumber\\
&+&\left(\alpha \ p_x + 2\sqrt{\frac{2}{3}} (1-3\delta)
 \ T_{xz} \cos\theta\right) \sin\theta \ \cos\phi \nonumber\\
&+&\left(\alpha \ p_y + 2\sqrt{\frac{2}{3}} (1-3\delta)
 \ T_{yz} \cos\theta\right) \sin\theta \ \sin\phi \nonumber\\
&+&(1-3\delta) \left(\frac{T_{xx}-T_{yy}}{\sqrt{6}} \right) \sin^2\theta
\cos(2\phi)\nonumber\\
&+&\left. \sqrt{\frac{2}{3}}(1-3\delta) \ T_{xy} \ \sin^2\theta \
\sin(2\phi) \right]. \label{angular_distribution}
\end{eqnarray}
Using partial integration of this differential distribution of the final 
state $f$ one can construct several asymmetries to probe various 
polarization parameters in Eq.~(\ref{Polarization_matrix}) which will be 
discussed in the next section.

\subsection{Estimation of polarization parameters}
For the process of interest, one might want to calculate the polarization
parameters of the particle $V$. This can be achieved at two levels: At production
process level and at the level of decay products. At the production level we 
first calculate the production density matrix,
Eq.~(\ref{production_density_matrix}), using helicity amplitudes of the
production process and then calculate the polarization matrix,
Eq.~(\ref{Polarization_matrix}). The polarization parameters can be extracted
from the polarization matrix elements as
\begin{eqnarray}\label{eq:pol_prod}
&&P_x= 
\frac{\bigg[[\rho_T(+,0)+\rho_T(0,+)]+[\rho_T(0,-)+\rho_T(-,0)] \bigg]}{\sqrt{2}\sigma_V},\nonumber\\
&&P_y =
 \frac{i\bigg[[\rho_T(0,+)-\rho_T(+,0)]+[\rho_T(-,0)-\rho_T(0,-)] \bigg]}{\sqrt{2}\sigma_V},\nonumber\\
&&P_z =
 \dfrac{\bigg[\rho_T(+,+)-\rho_T(-,-)\bigg]}{\sigma_V},\nonumber\\
&&T_{xy} =
 \frac{i\sqrt{6} \bigg[\rho_T(-,+)-\rho_T(+,-) \bigg]}{4\sigma_V},\nonumber\\
&&T_{xz} =
 \frac{\sqrt{3}\bigg[[\rho_T(+,0)+\rho_T(0,+)]-[\rho_T(0,-)+\rho_T(-,0)] \bigg]}{4\sigma_V},\nonumber\\
&&T_{yz} =
 \frac{i\sqrt{3}\bigg[[\rho_T(0,+)-\rho_T(+,0)]-[\rho_T(-,0)-\rho_T(0,-)] \bigg]}{4\sigma_V},\nonumber\\
&&T_{xx}-T_{yy} =
  \frac{\sqrt{6} \bigg[\rho_T(-,+)+\rho_T(+,-)\bigg]}{2\sigma_V},\nonumber\\
&&T_{zz}=
\dfrac{\sqrt{6}}{2}\bigg[\frac{\rho_T(+,+)+\rho_T(-,-)}{\sigma_V}-\frac{2}{3}\bigg],\nonumber\\
&&\hspace{0.5cm} = \dfrac{\sqrt{6}}{2}\bigg[\frac{1}{3}-\frac{\rho_T(0,0)}{\sigma_V}\bigg].
\end{eqnarray}
Using the tracelessness of $T_{ij}$, $T_{xx}+T_{yy}+T_{zz}=0$, along with 
values of $T_{zz}$ and $T_{xx}-T_{yy}$ from above, one can calculate 
$T_{xx}$ and $T_{yy}$ separately.

At the level of decay products, as in a real experiment or in a Monte-Carlo
simulator, one needs to compute specific asymmetries to extract the 
polarization parameters. 
For example, we can get $P_x$ from the left-right asymmetry as
\begin{eqnarray}\label{eq:pol_decay_Ax}
A_x&=&\frac{1}{\sigma}\bigg[\int _{\theta =0}^{\pi }\int _{\phi =
-\frac{\pi}{2}}^{\frac{\pi}{2}}  \dfrac{d\sigma}{d\Omega_f}
d\Omega_f -\int _{\theta=0}^{\pi }
\int _{\phi=\frac{\pi}{2}}^{3\frac{\pi}{2}}
\dfrac{d\sigma}{d\Omega_f}
d\Omega_f \bigg]\nonumber\\
&=&\frac{3 \alpha  P_x}{4}
\equiv  \dfrac{\sigma(\cos\phi>0)-\sigma(\cos\phi<0)}{\sigma(\cos\phi>0)+\sigma(\cos\phi<0)}.
\end{eqnarray}
\begin{figure*}
\centering
\includegraphics[width=7.43cm]{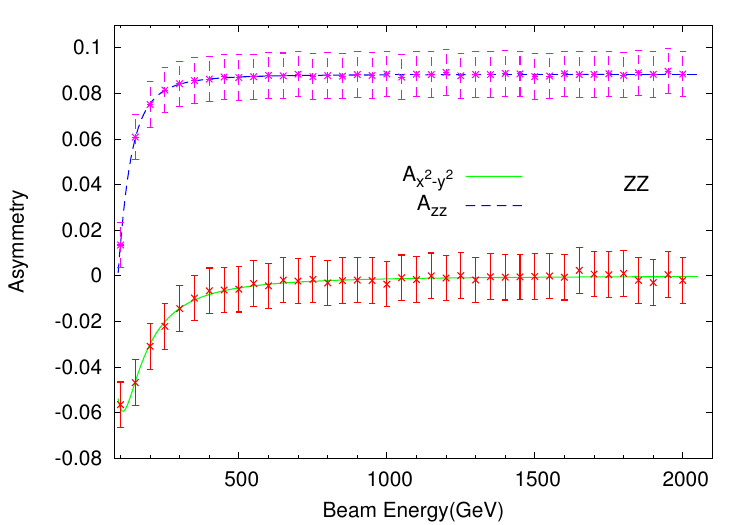}
\includegraphics[width=7.43cm]{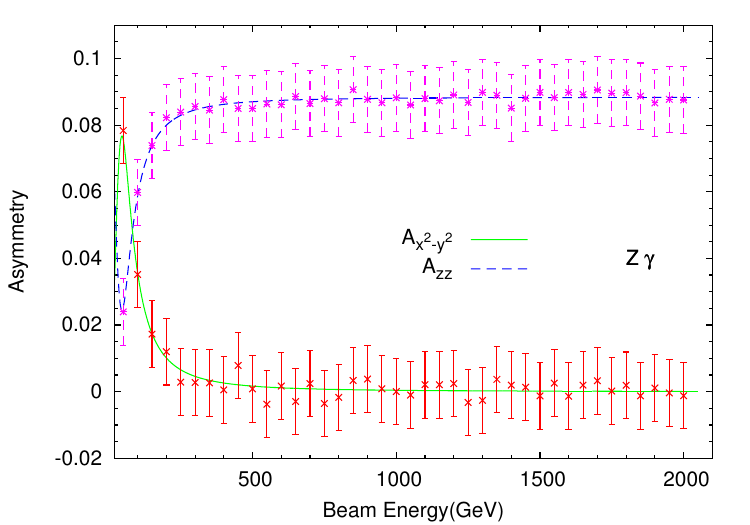}
\caption{\label{fig:asymmetry_sanity_check} The SM values (analytic) of asymmetries
$A_{x^2-y^2}$ ({\em solid/green line}) and $A_{zz}$ ({\em dashed/blue line}) as a function of beam
energy of the $e^+e^-$ collider for $ZZ$ ({\em left panel}) and $Z\gamma$ ({\em right panel}) 
processes. The data points with error bar corresponds to $10^4$ events
generated by {\tt MadGraph5}}
\end{figure*}
All other polarization parameters can  be obtained in a similar manner,
using Eq.~(\ref{angular_distribution}):
\begin{eqnarray}\label{eq:asymmetry_extraction_formulae}
A_y&\equiv & \dfrac{\sigma(\sin\phi>0)-\sigma(\sin\phi<0)}{\sigma(\sin\phi>0)+\sigma(\sin\phi<0)}
=\frac{3 \alpha  P_y}{4},\nonumber\\
A_z&\equiv & \dfrac{\sigma(\cos\theta>0)-\sigma(\cos\theta<0)}{\sigma(\cos\theta>0)+\sigma(\cos\theta<0)}
=\frac{3 \alpha  P_z}{4},\nonumber\\
A_{xy}&\equiv & \dfrac{\sigma(\sin 2\phi>0)-\sigma(\sin 2\phi<0)}{\sigma(\sin 2\phi>0)+\sigma(\sin 2\phi<0)}\nonumber\\
&=& \frac{2}{\pi } \sqrt{\frac{2}{3}} (1-3 \delta ) T_{xy},\nonumber\\
A_{xz}&\equiv & \dfrac{\sigma(\cos\theta\cos\phi<0)-\sigma(\cos\theta\cos\phi>0)}{\sigma(\cos\theta\cos\phi>0)+\sigma(\cos\theta\cos\phi<0)}\nonumber\\
&=&-\frac{2}{\pi } \sqrt{\frac{2}{3}} (1-3 \delta ) T_{xz},\nonumber\\
A_{yz}&\equiv & \dfrac{\sigma(\cos\theta\sin\phi>0)-\sigma(\cos\theta\sin\phi<0)}{\sigma(\cos\theta\sin\phi>0)+\sigma(\cos\theta\sin\phi<0)}\nonumber\\
&=& \frac{2 }{\pi }\sqrt{\frac{2}{3}} (1-3 \delta ) T_{\text{yz}},\nonumber\\
A_{x^2-y^2}&\equiv & \dfrac{\sigma(\cos 2\phi>0)-\sigma(\cos 2\phi<0)}{\sigma(\cos 2\phi>0)+\sigma(\cos 2\phi<0)}\nonumber\\
 &=& \frac{1}{\pi }\sqrt{\frac{2}{3}} (1-3 \delta ) \left(T_{xx}-T_{yy}\right),\nonumber\\
A_{zz}&\equiv & \dfrac{\sigma(\sin 3\theta>0)-\sigma(\sin 3\theta<0)}{\sigma(\sin 3\theta>0)+\sigma(\sin 3\theta<0)}\nonumber\\
 &=& \frac{3}{8}\sqrt{\frac{3}{2}} (1-3 \delta ) T_{zz}.
\end{eqnarray}
We note that pure azimuthal asymmetries ($A_x$, $A_y$, 
$A_{xy}$, $A_{x^2-y^2}$) listed above were already discussed in 
Ref.~\cite{Boudjema:2009fz},
however, a complete set of asymmetries for the $W^\pm$-boson ($\delta=0,~\alpha=-1$) 
is given in Ref.~\cite{Aguilar-Saavedra:2015yza}. 

While extracting the polarization asymmetries in the collider/event generator 
we have to make sure that the analysis is done in the rest frame of $V$. The
initial beam defines the $z$-axis in the lab, while the production plane of $V$
defines the $xz$ plane, i.e. $\phi=0$ plane. While boosting to the rest frame
of $V$ we keep the $xz$ plane invariant. The polar and the azimuthal angles of
the decay products of $V$ are measured with respect to the {\em would-be}
momentum of $V$.

As a demonstration of the two methods mentioned above, we look at two processes:
$e^+e^-\to~ZZ$ and $e^+e^-\to~Z\gamma$. The polarization parameters are
constructed both at the production level, using Eq.~(\ref{eq:pol_prod}), and
at the decay level, using Eqs.~(\ref{eq:pol_decay_Ax}) and
(\ref{eq:asymmetry_extraction_formulae}). We observe that out of eight
polarization asymmetries only three, $A_x$, $A_{x^2-y^2}$, and $A_{zz}$, are
non-zero in the SM. The asymmetries $A_{x^2-y^2}$ and $A_{zz}$ are calculated
analytically for the production part and shown as a function of beam energy in 
Fig.~\ref{fig:asymmetry_sanity_check} with solid lines.
For the same processes with $Z\to f\bar f$ decay, we generate events
using {\tt MadGraph5}~\cite{Alwall:2014hca} with different values of beam 
energies. The
polarization asymmetries were constructed from these events and are shown in
Fig.~\ref{fig:asymmetry_sanity_check} with points. The statistical error 
bars shown correspond to $10^4$ events. We observe  agreement
between the asymmetries calculated at the production level (analytically) and
the decay level (using event generator). 
Any possible new physics in the production process of $Z$ boson
is expected to change the cross section, kinematical distributions and the
values of the polarization parameters/asymmetries. We intend to use these
asymmetries to probe the standard and BSM physics.

\section{Anomalous Lagrangian and their probes}
\label{sec:Lag}
The effective Lagrangian for the anomalous trilinear gauge boson interactions 
in the neutral sector is given in Eq.~(\ref{LaTGCfull}), which includes both 
dimension-$6$ and dimension-$8$ operators as found in the literature. 
For the present work 
we restrict our analysis to dimension-$6$ operators only. Thus, the anomalous 
Lagrangian of our interest is
\begin{eqnarray}\label{aTGC_Lagrangian}
{\cal L}_{aTGC}^{dim-6}&=&
 \frac{g_e}{M_Z^2} \Bigg [
-[f_4^\gamma (\partial_\mu F^{\mu \beta})+
f_4^Z (\partial_\mu Z^{\mu \beta}) ] Z_\alpha 
( \partial^\alpha Z_\beta)\nonumber\\
&+&[f_5^\gamma (\partial^\sigma F_{\sigma \mu})+
f_5^Z (\partial^\sigma Z_{\sigma \mu}) ] \wtil{Z}^{\mu \beta} Z_\beta\nonumber \\
&-&  [h_1^\gamma (\partial^\sigma F_{\sigma \mu})
+h_1^Z (\partial^\sigma Z_{\sigma \mu})] Z_\beta F^{\mu \beta}\nonumber\\
&-&[h_3^\gamma  (\partial_\sigma F^{\sigma \rho})
+ h_3^Z  (\partial_\sigma Z^{\sigma \rho})] Z^\alpha
 \wtil{F}_{\rho \alpha}
\Bigg ]. 
\end{eqnarray}
This yields anomalous vertices $ZZZ$ through $f^Z_{4,5}$ couplings, $\gamma ZZ$ 
through $f^\gamma_{4,5}$ and $h^Z_{1,3}$ couplings and $\gamma\gamma Z$ through
$h^\gamma_{1,3}$ couplings. There is no $\gamma\gamma\gamma$ vertex in the above
Lagrangian.
\begin{figure}[h!]
\includegraphics[width=8.0cm]{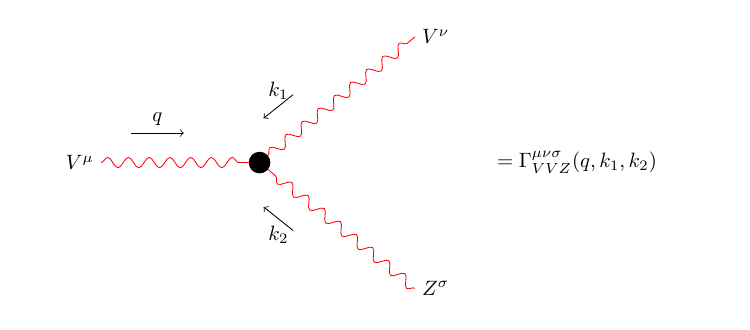}
\caption{\label{fig:aTGC_off-shell_Vertex} Feynman diagram for a general
anomalous triple gauge boson vertex with  $V=Z/\gamma$}
\end{figure}
We use {\tt FeynRules}~\cite{Alloul:2013bka} to obtain the vertex tensors and 
they are given by
\begin{eqnarray}\label{eq:aTGC_off-shel_vertex_azz}
&&\Gamma_{\gamma Z Z}^{\mu\nu\sigma}(q,k_1,k_2)=\nonumber\\
&&\dfrac{g_e}{M_Z^2}\Bigg[f_4^\gamma \bigg(\left(k_2^{\nu}g^{\mu\sigma} +k_1^{\sigma} g^{\mu\nu}\right) q^2 -q^{\mu} \left(k_1^{\sigma} q^{\nu}+k_2^{\nu} q^{\sigma}\right)\bigg)\Bigg.\nonumber\\
&&+f_5^\gamma \left(q^{\mu}{q}_{\beta} \epsilon ^{\nu\sigma\alpha\beta} +q^2 \epsilon ^{\mu\nu\sigma\alpha}\right) (k_1-k_2)_{\alpha}\nonumber\\ 
&&+h_1^Z \bigg( k_2^{\mu}  q^{\nu} k_2^{\sigma} +  k_1^{\mu} k_1^{\nu} q^{\sigma} +\left(k_1^2-k_2^2\right) \left(q^{\nu}g^{\mu\sigma}-q^{\sigma}g^{\mu\nu}\right)\bigg.\nonumber\\
&&\bigg.-k_2^{\sigma} g^{\mu\nu} q.k_2-k_1^{\nu} g^{\mu\sigma}q.k_1  \bigg) \nonumber\\ 
&& \Bigg. - h_3^Z \left( k_1^{\nu} {k_1}_{\beta}\epsilon ^{\mu\sigma\alpha\beta} +{k_2}_{\beta} k_2^{\sigma} \epsilon ^{\mu\nu\alpha\beta}
+\left(k_2^2-k_1^2\right) \epsilon ^{\mu\nu\sigma\alpha}\right) {q}_{\alpha}\Bigg],\nonumber\\
\end{eqnarray}
\begin{eqnarray}\label{eq:aTGC_off-shel_vertex_zzz}
&&\Gamma_{Z Z Z}^{\mu\nu\sigma}(q,k_1,k_2)=\nonumber\\
&&\dfrac{g_e}{M_Z^2}\Bigg[f_4^Z\bigg(-q^{\mu} q^{\nu} k_1^{\sigma} -k_2^{\mu} q^{\nu} k_2^{\sigma} - k_2^{\mu}k_1^{\nu} k_1^{\sigma} -k_1^{\mu} k_2^{\nu} k_2^{\sigma}\bigg.\Bigg.\nonumber\\
&&-\left( q^{\mu} k_2^{\nu} +k_1^{\mu} k_1^{\nu} \right) q^{\sigma}
 + g^{\mu\nu}\left(q^2 k_1^{\sigma} +k_1^2 q^{\sigma} \right)\nonumber\\
&&\bigg. + g^{\mu\sigma}\left(q^2 k_2^{\nu} +k_2^2 q^{\nu} \right)+g^{\nu\sigma}\left(k_2^2 k_1^{\mu} +k_1^2 k_2^{\mu} \right)\bigg)\nonumber\\
&&- f_5^Z\bigg(\epsilon ^{\mu\nu\alpha\beta}(k_1-q)_{\alpha} {k_2}_{\beta} k_2^{\sigma}+\epsilon ^{\mu\nu\sigma\alpha}\left(\left(k_1^2-k_2^2\right){q}_{\alpha}\right.\bigg.\nonumber\\
&&\left.+\left(k_2^2-q^2\right){k_1}_{\alpha}
+\left(q^2-k_1^2\right){k_2}_{\alpha}\right)\nonumber\\
&&\Bigg.\bigg.+{k_1}_{\beta} k_1^{\nu}(k_2-q)_{\alpha}\epsilon ^{\mu\sigma\alpha\beta}+{q}_{\beta} q^{\mu}(k_2-k_1)_{\alpha}\epsilon ^{\nu\sigma\alpha\beta}\bigg)\Bigg]\nonumber ,\\ 
\end{eqnarray}
\begin{eqnarray}\label{eq:aTGC_off-shel_vertex_aaz} 
&& \Gamma_{\gamma \gamma Z}^{\mu\nu\sigma}(q,k_1,k_2)=\nonumber\\
&&\dfrac{g_e}{M_Z^2}\Bigg[ h_1^\gamma \bigg(q^{\mu}q^{\nu} k_1^{\sigma}+q^{\sigma}k_1^{\mu} k_1^{\nu}-g^{\mu\nu}\left(q^2 k_1^{\sigma}+k_1^2 q^{\sigma} \right)\Bigg.\bigg.\nonumber\\
&&\bigg.+ g^{\mu\sigma}\left(k_1^2 q^{\nu}-q.k_1 k_1^{\nu} \right)  + g^{\nu\sigma}\left(q^2 k_1^{\mu}-q.k_1 q^{\mu} \right)\bigg)\nonumber\\
&& - h_3^\gamma \bigg({k_1}_{\beta} k_1^{\nu}{q}_{\alpha} \epsilon ^{\mu\sigma\alpha\beta} +q^{\mu} {k_1}_{\alpha}{q}_{\beta} \epsilon ^{\nu\sigma\alpha\beta} \bigg.\nonumber\\
&&+\Bigg.\bigg.\left(q^2 {k_1}_{\alpha} - k_1^2 {q}_{\alpha} \right) \epsilon ^{\mu\nu\sigma\alpha}\bigg)\Bigg].
\end{eqnarray}

The notations for momentum and Lorentz indices are shown in  
Fig.~\ref{fig:aTGC_off-shell_Vertex}. We are interested in possible trilinear
gauge boson vertices appearing in the processes $e^+e^-\to ZZ$ and 
$e^+e^-\to Z\gamma$ with final state gauge bosons being on-shell. 
For the process $e^+e^-\to ZZ$, the vertices $\gamma^\star Z Z$ and
$Z^\star Z Z$ appear with on-shell conditions $k_1^2=k_2^2=M_Z^2$. The terms 
proportional to $k_1^\nu$ and $k_2^\sigma$ in 
Eqs.~(\ref{eq:aTGC_off-shel_vertex_azz}) and 
(\ref{eq:aTGC_off-shel_vertex_zzz})
vanish due to the transversity of the polarization states. Thus, in the 
on-shell case, the vertices for $e^+e^-\to ZZ$ reduce to
\begin{eqnarray}\label{eq:aTGC_vertex_zz}
\Gamma_{V^\star Z Z}^{\mu\nu\sigma}(q,k_1,k_2)=&&
 -\dfrac{g_e}{M_Z^2}\left(q^2 - M_V^2 \right)\Bigg[f_4^V \left(q^{\sigma} g^{\mu\nu} +q^{\nu} g^{\mu\sigma} \right)\Bigg.\nonumber\\
  &&\Bigg.- f_5^V  \epsilon _{}^{\mu\nu\sigma\alpha}  (k_1 - k_2)_{\alpha}\Bigg].
\end{eqnarray}
For the process $e^+e^-\to Z\gamma$  the vertices $\gamma Z^\star Z$ and
$\gamma^\star \gamma Z$  appear with corresponding on-shell and polarization
transversity conditions. Putting these conditions in 
Eqs.~(\ref{eq:aTGC_off-shel_vertex_azz}) and 
(\ref{eq:aTGC_off-shel_vertex_aaz}) and some relabelling of momenta etc. in
Eq.~(\ref{eq:aTGC_off-shel_vertex_azz}) the relevant vertices  $Z^\star \gamma
Z$ and  $\gamma^\star \gamma Z$ can be represented together by
\begin{eqnarray}\label{eq:aTGC_vertex_za}
\Gamma_{V^\star\gamma  Z}^{\mu\nu\sigma}(q,k_1,k_2)=&&
 \dfrac{g_e}{M_Z^2}\left(q^2 - M_V^2 \right)\Bigg[h_1^V  \left(k_1^{\mu} g^{\nu\sigma} - k_1^{\sigma} g^{\mu\nu} \right)\nonumber\\
&&- h_3^V   \epsilon _{}^{\mu\nu\sigma\alpha}{k_1}_{\alpha}\Bigg].
\end{eqnarray}
The off-shell $V^\star$ is the propagator in our processes and 
couples to the massless electron current, as shown in 
Fig.~\ref{fig:Z_boson_production_Feynman_diagram}(c), (d).  After 
above-mentioned reduction of the vertices, there were some terms proportional to
$q^\mu$ that yield zero upon contraction with the electron current, hence they are 
dropped from the above expressions.
We note that although $h^Z$ and $f^\gamma$ appear together in the off-shell 
vertex of $\gamma ZZ$ in Eq.~(\ref{eq:aTGC_off-shel_vertex_azz}), they decouple 
after choosing separate processes; the $f^V$ appear only in $e^+e^-\to ZZ$,
while the $h^V$ appear only in $e^+e^-\to Z\gamma$. This decoupling simplifies our
analysis as we can study two processes independent of each other when 
we perform a global fit to the parameters in Section~\ref{sec:MCMC}.

\section{Asymmetries, limits, and sensitivity to anomalous couplings}
\label{sec:asym_sen}
\begin{figure*}
\centering
\includegraphics[width=16.0cm]{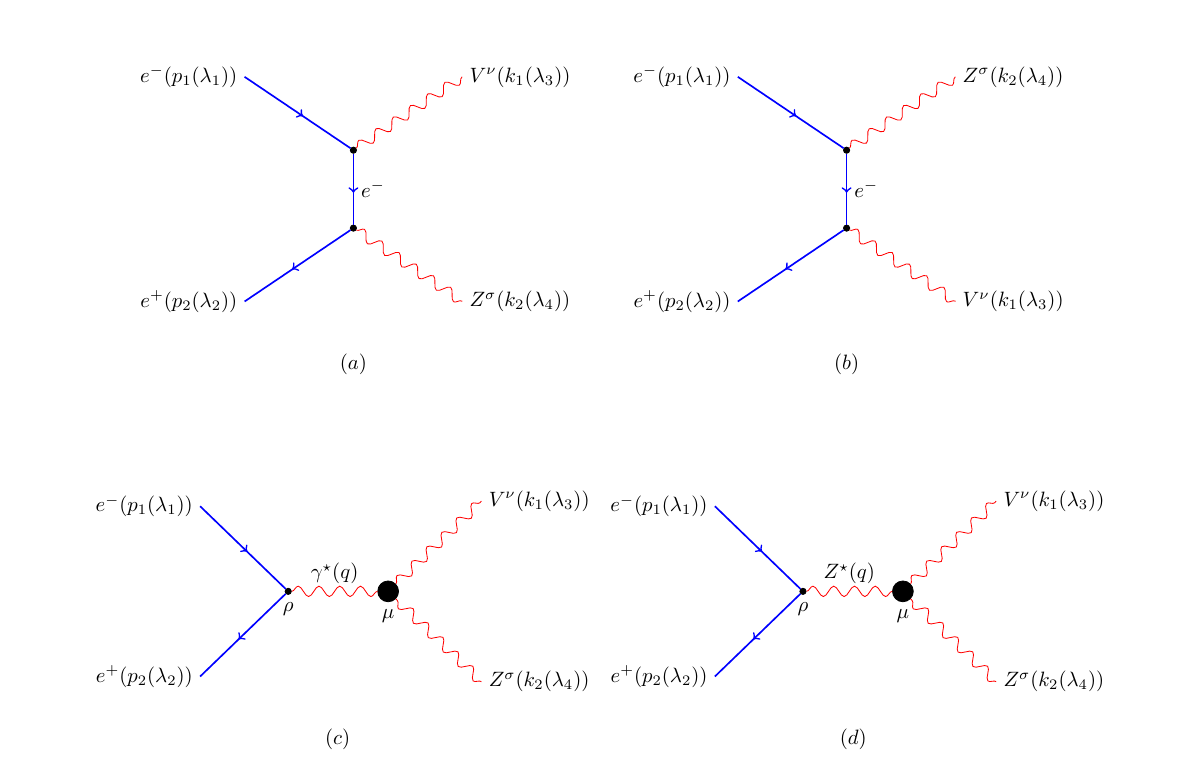}
\caption{\label{fig:Z_boson_production_Feynman_diagram} Feynman
diagrams for the production of $ZZ$ or $Z\gamma$ at 
$e^+e^-$ collider }
\end{figure*}
In this section we thoroughly investigate the effects of anomalous couplings
in the processes $e^+e^-\to ZZ$ and $e^+e^-\to Z\gamma$. 
We use tree level SM interactions along-with anomalous couplings 
shown in Eq.~(\ref{aTGC_Lagrangian}) for our analysis.
The Feynman diagrams for these processes are  given in the 
Fig.~\ref{fig:Z_boson_production_Feynman_diagram} where the anomalous 
vertices are 
shown as big blobs. The helicity amplitudes for the anomalous part together with SM
contributions for both these processes are given in~\ref{appendix:helicity_amplitude}. 
These are then used to calculate  the polarization observables and the cross section which are given 
in~\ref{appendix:Expresson_observables}.

\subsection{Parametric dependence of asymmetries on anomalous couplings}
The dependences of the observables on the anomalous couplings for the $ZZ$ and 
$Z\gamma$ processes are given in Tables~\ref{tab:parameter_dependent_pol_zz} and~\ref{tab:parameter_dependent_pol_za}, respectively.

\begin{table}
\centering
\caption{\label{tab:parameter_dependent_pol_zz} Dependence of the polarization 
observables on the anomalous coupling in $ZZ$ final state}
\renewcommand{\arraystretch}{1.5}
\begin{tabular*}{\columnwidth}{@{\extracolsep{\fill}}lll@{}}
\hline
Observables & Linear terms & Quadratic terms\\ \hline
$ \sigma $& $f_5^Z,f_5^\gamma$ & $(f_4^\gamma)^2, (f_5^\gamma)^2, (f_4^Z)^2, 
(f_5^Z)^2, f_4^\gamma f_4^Z, f_5^\gamma f_5^Z $  \\ 
$\sigma \times A_x$& $ f_5^\gamma ,f_5^Z $ & $-$\\ 
$\sigma \times A_y $& $f_4^\gamma ,f_4^Z $ & $-$ \\ 
$\sigma \times A_{xy} $& $f_4^Z,f_4^\gamma$ & $f_4^Zf_5^\gamma ,f_4^\gamma
f_5^Z,f_4^\gamma f_5^\gamma,f_4^Zf_5^Z $ \\ 
$\sigma\times A_{x^2-y^2}$& $f_5^Z,f_5^\gamma$ & 
$(f_4^\gamma)^2, (f_5^\gamma)^2, (f_4^Z)^2, 
(f_5^Z)^2, f_4^\gamma f_4^Z, f_5^\gamma f_5^Z $  \\
$\sigma\times A_{zz} $& $f_5^Z,f_5^\gamma$ & 
$(f_4^\gamma)^2, (f_5^\gamma)^2, (f_4^Z)^2, 
(f_5^Z)^2, f_4^\gamma f_4^Z, f_5^\gamma f_5^Z $  \\  \hline
\end{tabular*}
\end{table}
\begin{table}
\centering
\caption{\label{tab:parameter_dependent_pol_za} Dependence of the polarization 
observables on the anomalous coupling in $Z\gamma$ final state}
\renewcommand{\arraystretch}{1.5}
\begin{tabular*}{\columnwidth}{@{\extracolsep{\fill}}lll@{}}\hline
Observables & Linear terms & Quadratic terms\\ \hline
$ \sigma $& $h_3^Z,h_3^\gamma$ & $(h_1^\gamma)^2,(h_3^\gamma)^2,(h_1^Z)^2,(h_3^Z)^2,h_1^\gamma h_1^Z,h_3^\gamma h_3^Z $  \\ 
$\sigma \times A_x$&  $h_3^Z,h_3^\gamma$ & $(h_1^\gamma)^2,(h_3^\gamma)^2,(h_1^Z)^2,(h_3^Z)^2,h_1^\gamma h_1^Z,h_3^\gamma h_3^Z $ \\ 
$\sigma \times A_y $& $h_1^\gamma ,h_1^Z $ & $-$ \\ 
$\sigma \times A_{xy} $& $h_1^\gamma ,h_1^Z $ & $-$ \\ 
$\sigma\times A_{x^2-y^2}$& $h_3^\gamma ,h_3^Z $ & 
$- $  \\ 
$\sigma\times A_{zz} $&$h_3^Z,h_3^\gamma$ & $(h_1^\gamma)^2,(h_3^\gamma)^2,(h_1^Z)^2,(h_3^Z)^2,h_1^\gamma h_1^Z,h_3^\gamma h_3^Z $   \\ \hline
\end{tabular*}
\end{table}
In the SM, the helicity amplitudes are real, 
thus the production density matrix elements in 
Eq.~(\ref{production_density_matrix}) are all  real. This implies  
$A_y$, $A_{xy}$ and $A_{yz}$ are all zero in the SM: see 
Eq.~(\ref{eq:pol_prod}). The asymmetries $A_z$ and $A_{xz}$ are also zero 
for the SM couplings due to the forward-backward symmetry of the $Z$ boson in the 
c.m. frame, owing to the presence of both $t$- and $u$-channel
diagrams and unpolarized initial beams.
After including anomalous couplings, $A_{y}$ and $A_{xy}$ receive a 
non-zero contribution, while $A_z$, $A_{xz}$ and $A_{yz}$ 
remain zero for the unpolarized initial beams.

From the list of non-vanishing asymmetries, only $A_y$ and $A_{xy}$ are CP-odd,
while the others
are CP-even. All the CP-odd observables are linearly dependent upon the CP-odd
couplings, like $f_4^V$ and $h_1^V$, while all the CP-even observables have
only quadratic dependence on the CP-odd couplings. In the SM, the $Z$ boson's
couplings respect CP symmetry; thus $A_y$ and $A_{xy}$ vanish. 
Hence, any significant deviation of $A_{y}$ and $A_{xy}$ 
from zero at the collider will indicate a clear sign of CP-violating new 
physics interactions. Observables that have only a linear dependence on the
anomalous couplings yield a {\it single interval limits} on these couplings.
On the other hand, any quadratic appearance (like $(f_5^V)^2$ in $\sigma$) may
yield more than one interval of the couplings, while putting limits. For the 
case of $ZZ$ process, $A_x$ and $A_y$ do not have any quadratic dependence, hence they 
yield the cleanest limits on the CP-even and -odd parameters, respectively.
Similarly, for the $Z \gamma $ process we have $A_y$, $A_{xy}$, and $A_{x^2-y^2}$, 
which have only a linear dependence and provide clean limits. 
These clean limits, however, may not be the strongest limits as we will see 
in the following sections. 

\subsection{Sensitivity and limits on anomalous couplings}
Sensitivity of an observable $\mathcal{O}$ dependent on parameter $\vec{f}$ is
defined as
\begin{equation}\label{eq:sensitivity}
{\cal S}(\mathcal{O}(\vec{f}))=\dfrac{|\mathcal{O}(\vec{f})-\mathcal{O}(\vec{f}=0)|}{\delta \mathcal{O}},
\end{equation}
where $\delta \mathcal{O}=\sqrt{(\delta \mathcal{O}_{stat.})^2 + (\delta
\mathcal{O}_{sys.})^2}$ is the estimated error in $\mathcal{O}$. 
If the observable is an asymmetry, $A=(N^+ - N^-)/(N^+ + N^-)$, the 
error is given by 
\begin{equation}\label{eq:error_in_asymmetry}
\delta A= \sqrt{\frac{1-A^2}{{\cal L}\sigma}+\epsilon_A^2},
\end{equation}
where $N^+ + N^-=N_T=L\sigma$, ${\cal L}$ being the integrated luminosity of the 
collider.  The error in the cross section $\sigma$ will be given by
\begin{equation}\label{eq:error_in_sigma}
\delta\sigma= \sqrt{\frac{\sigma}{{\cal L}} + (\epsilon \sigma)^2}. 
\end{equation}
Here $\epsilon_A$ and $\epsilon$ are the systematic fractional errors in $A$ and 
$\sigma$,  respectively, while remaining one are statistical errors.
\begin{figure*}
\centering
\includegraphics[width=7.43cm]{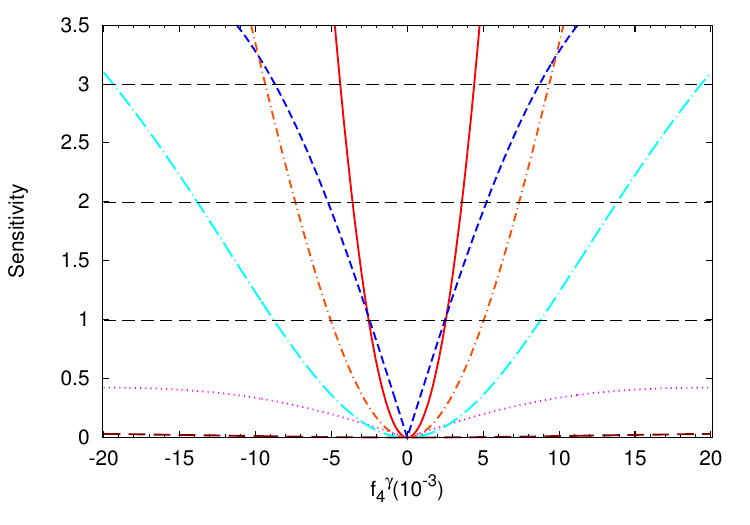}
\includegraphics[width=7.43cm]{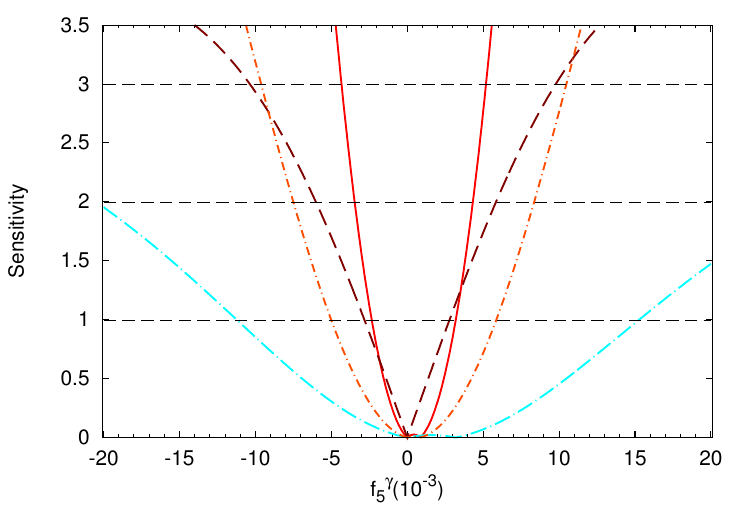}
\includegraphics[width=7.43cm]{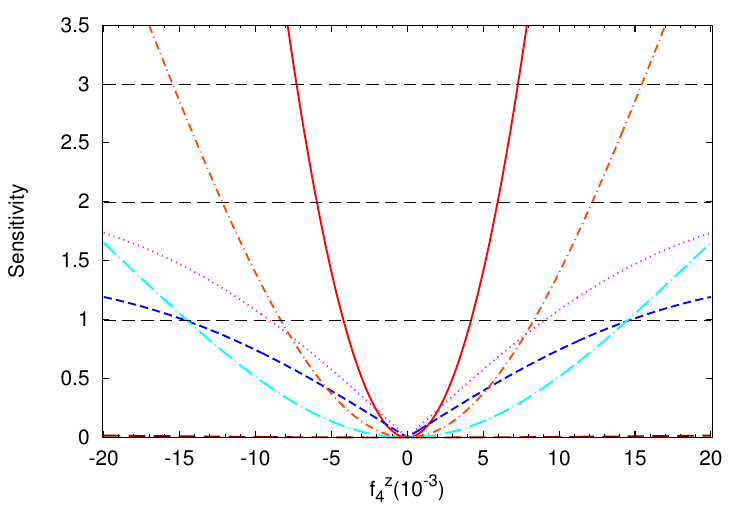}
\includegraphics[width=7.43cm]{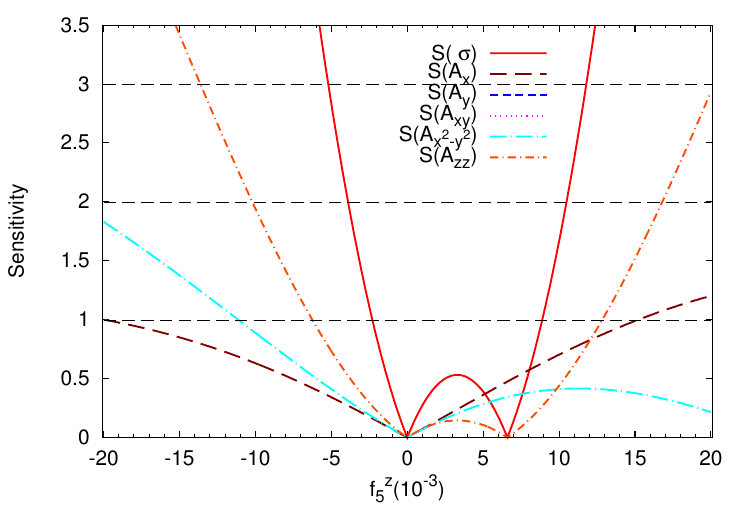}
\caption{\label{fig:one_parameter_sensitivity_zz} Sensitivity of the cross section
and asymmetries to anomalous couplings for the process $e^+e^-\rightarrow ZZ$
with $\sqrt{\hat{s}}=500$ GeV and $L=100$ fb$^{-1}$ }
\end{figure*}
\begin{figure*}
\centering
\includegraphics[width=7.43cm]{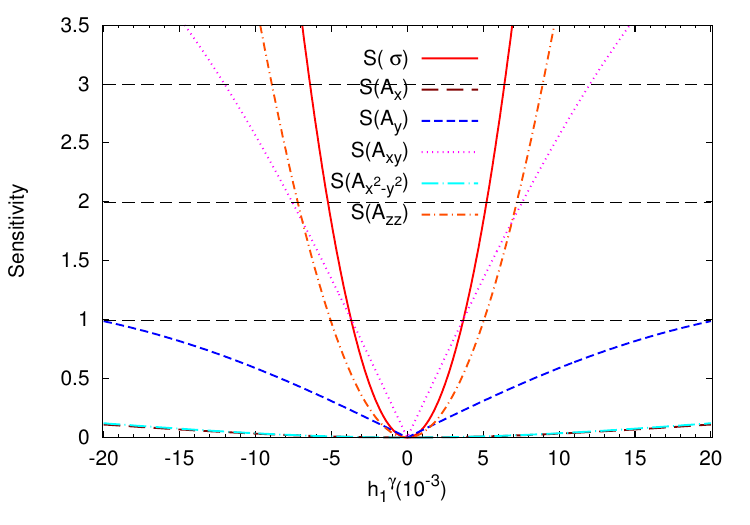}
\includegraphics[width=7.43cm]{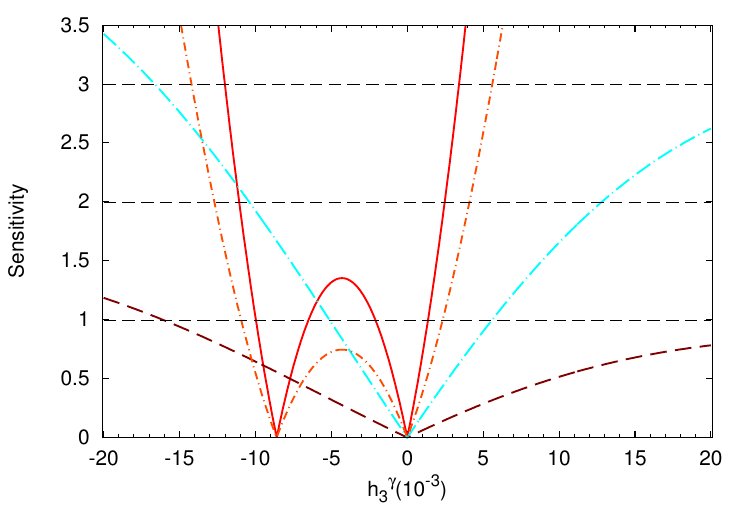}
\includegraphics[width=7.43cm]{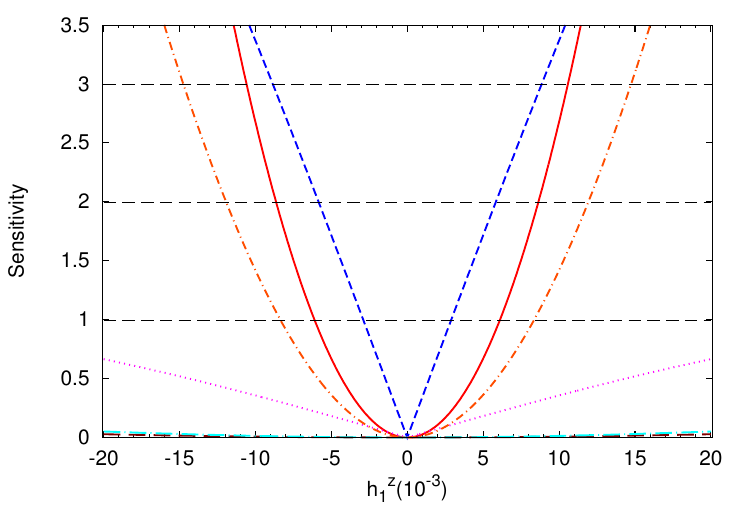}
\includegraphics[width=7.43cm]{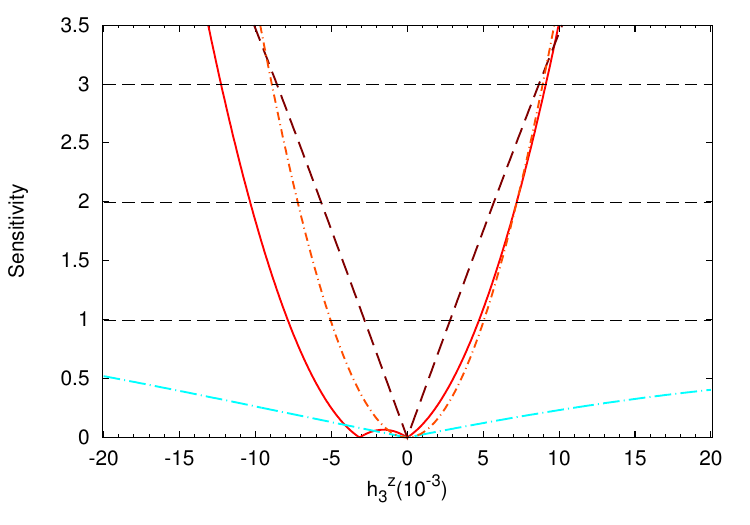}
\caption{\label{fig:one_parameter_sensitivity_za} Sensitivity of the cross section
and asymmetries to anomalous couplings for the process 
$e^+e^-\rightarrow Z\gamma$ with $\sqrt{\hat{s}}=500$ 
GeV, $L=100$ fb$^{-1}$, and $10^\circ \le\theta_\gamma\le170^\circ$}
\end{figure*}

For numerical calculations, we choose ILC running at c.m. energy 
$\sqrt{\hat{s}}=500$ GeV and integrated luminosity ${\cal L}=100$ fb$^{-1}$.
We use $\epsilon_A=\epsilon=0$ for most of our analysis, however,
 we do discuss the impact of systematic errors on our results.
 With this
choice the sensitivity of all the polarization asymmetries, 
Eq.~(\ref{eq:asymmetry_extraction_formulae}), and the cross section have been
calculated varying one parameter at a time. These sensitivities are shown
in Figs.~\ref{fig:one_parameter_sensitivity_zz} and~\ref{fig:one_parameter_sensitivity_za} 
for the $ZZ$ and $Z\gamma$ processes, respectively, for each observable.
In the $e^+e^-\to Z\gamma$ process we have taken a cut-off on the polar angle,
$10^\circ \le\theta_\gamma\le170^\circ$ to keep away from the beam pipe.  
For these limits the analytical expressions  shown in~\ref{appendix:Expresson_observables}  are used.

We see that in the $ZZ$ process the tightest constraint on 
$f_4^\gamma$ at $1\sigma$ level comes from the asymmetry $A_{y}$ owing to 
its linear and strong dependence on the coupling. For $f_5^\gamma$, 
both $A_x$ and the cross section $\sigma_{ZZ}$, give comparable limits at 
$1\sigma$ but $\sigma_{ZZ}$ 
gives a tighter limit at higher values of sensitivity. This is because the
quadratic term in  $\sigma_{ZZ}$ comes with higher power of 
energy/momenta and
hence a larger sensitivity. Similarly, the strongest limit on 
$f_4^Z$ and $f_5^Z$ as well comes from $\sigma_{ZZ}$. Though the cross section
gives the tightest constrain on most of the coupling in $ZZ$ process, our 
polarization asymmetries also provide comparable limits. 
Another noticeable fact is that $\sigma_{ZZ}$ has a linear as well as quadratic 
dependence on $f_5^Z$ and the sensitivity curve is symmetric about a point 
larger than zero. 
Thus, when we do a parameter estimation exercise, we will always have 
a bias toward a positive value of $f_5^Z$. The same is the case with the coupling 
$f_5^\gamma$ , but the strength of the linear term is small and 
the sensitivity plot with $\sigma_{ZZ}$ looks almost symmetric about 
$f_5^\gamma=0$.

In the $Z\gamma$ process, the tightest constraint on $h_1^\gamma$ comes from 
$A_{xy}$, on $h_3^\gamma$ it comes from $\sigma_{Z \gamma }$, on it $h_1^Z$ comes 
from $A_y$, and on $h_3^Z$ it comes from $A_x$. The cross section $\sigma_{Z \gamma}$
and $A_{zz}$ has a linear as well as quadratic dependence on $h_3^\gamma$,
and $\sigma_{Z \gamma}$ and they give two intervals at $1\sigma$ level. Other 
observables can help resolve the 
degeneracy when we use more than one observables at a time. Still, the
cross section prefers a negative value of $h_3^\gamma$ and it will be seen again
in a multi-variate analysis.
The coupling $h_3^Z$ also has quadratic appearance in the cross section and 
yields a bias toward negative values of $h_3^Z$.

\begin{table*}
\centering
\caption{\label{tab:aTGC_constrain_form_1sigma_sensitivity} List of tightest 
limits on anomalous couplings at $1~\sigma$ level and the corresponding
observable obtained for $\sqrt{\hat{s}}=500$ GeV and $L=100$ fb$^{-1}$}
\renewcommand{\arraystretch}{1.5}
\begin{tabular*}{\textwidth}{@{\extracolsep{\fill}}llllll@{}}\hline
\multicolumn{3}{c}{$ZZ$ process} &
\multicolumn{3}{c}{$Z \gamma$ process}\\ \hline
Coupling & Limits & Comes from &
Coupling & Limits & Comes from\\ \hline
$|f_4^\gamma|$ & $ \le 2.4 \times 10^{-3}$ & $A_y$ & 
$|h_1^\gamma|$  & $\le 3.6\times 10^{-3}$ & $A_{xy}$, $\sigma$ \\ 
$|f_4^Z|$ &  $\le 4.2\times 10^{-3}$ &  $\sigma$ & 
$|h_1^Z|$ & $\le2.9\times 10^{-3}$ & $A_y$   \\ 
$f_5^\gamma$ & $\in[-2.3, 2.7] \times 10^{-3}$& $A_x$,
$\sigma$ & 
$h_3^\gamma$  & $\in [-2.1, 1.3] \times 10^{-3} $ &  $\sigma$ \\ 
 & & & &or $\in[-9.9, -6.5] \times 10^{-3} $& \\ 

$f_5^Z$ & $\in [-2.3, 8.8] \times 10^{-3}$ & $\sigma$ & 
$|h_3^Z|$ & $\le 2.8\times 10^{-3}$ & $A_x$ \\ \hline
\end{tabular*}
\end{table*}
The tightest limits on the anomalous couplings (at $1\sigma$ level), obtained
using one observable at a time and varying one coupling at a time, are listed in 
Table~\ref{tab:aTGC_constrain_form_1sigma_sensitivity}  
along with the corresponding observables. 
A comparison between 
Tables~\ref{tab:aTGC_constrain_form_collider} and~\ref{tab:aTGC_constrain_form_1sigma_sensitivity} shows that 
an $e^+e^-$ collider running at $500$ GeV and $100$ fb$^{-1}$ provides
better limits on the anomalous coupling ($f_i^V$) in the $ZZ$ process than the $7$ 
TeV LHC at $5$ fb$^{-1}$. For the $Z\gamma$ process the experimental limits are
available from $8$ TeV LHC with $19.6$ fb$^{-1}$ luminosity 
(Table~\ref{tab:aTGC_constrain_form_collider}) and they are comparable
to the single observable limits shown in
Table~\ref{tab:aTGC_constrain_form_1sigma_sensitivity}. 
These limits can be further improved if we use all the
observables in a $\chi^2$ kind of analysis.

We can further see that the sensitivity curves for CP-odd observables, $A_y$ 
and $A_{xy}$, has no or a very mild dependence on the CP-even couplings. The
mild dependence comes through the cross section $\sigma$,  sitting in 
the denominator of the asymmetries. We see that CP-even observables provide 
tight constraint on CP-even couplings and
CP-odd observables provide tight constraint on the CP-odd couplings. Thus, not
only  can we study the two processes independently, it is possible to study the
CP-even and CP-odd couplings almost independent of each other.
To this end, we shall perform a two-parameter sensitivity analysis in the next
subsection.

A note on the systematic error is in order. The sensitivity of an observable is
inversely  proportional to the size of its estimated error, 
Eq.~(\ref{eq:sensitivity}). Including the systematic error will
increase the size of the estimated error and hence decrease the
sensitivity. For example, including $\epsilon_A=1~\%$ to $L=100$ fb$^{-1}$ 
increases $\delta A$ by a factor of $1.3$ and dilutes the sensitivity by the 
same factor. This modifies the best limit on $|f_4^\gamma|$, coming from
$A_y$, to $2.97\times 10^{-3}$ (dilution by a factor of $1.3$); see
Table~\ref{tab:aTGC_constrain_form_1sigma_sensitivity}. 
For the cross section, adding $\epsilon=2~\%$ systematic error increases
$\delta\sigma$ by a factor of $1.5$. The best limit on $|f_4^Z|$, coming 
from the cross section, changes to $5.35\times 10^{-3}$, a dilution by a factor
of $1.2$. Since inclusion of the above systematic errors modifies the limits
on the couplings only by $20~\%$ to $30~\%$, we shall restrict ourselves to the 
statistical error for simplicity for rest of the analysis.

\subsection{Two-parameter sensitivity analysis}
\begin{figure*}
\centering
\includegraphics[width=4.9cm]{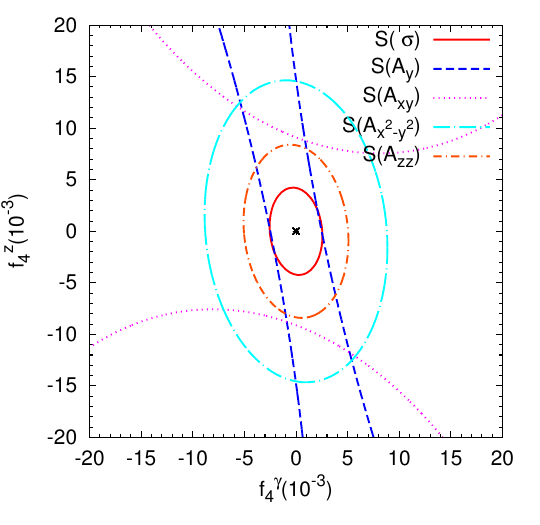}
\includegraphics[width=4.9cm]{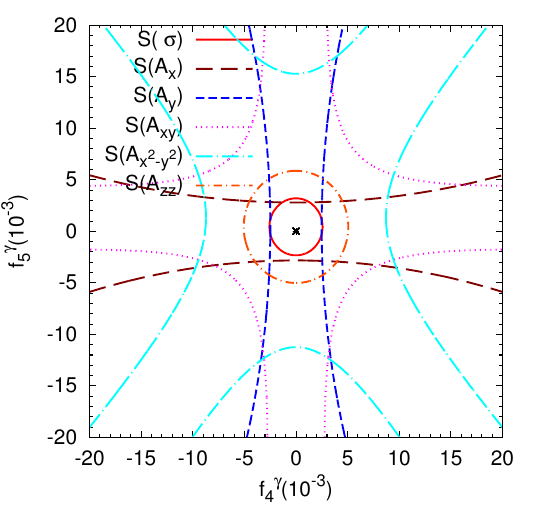}
\includegraphics[width=4.9cm]{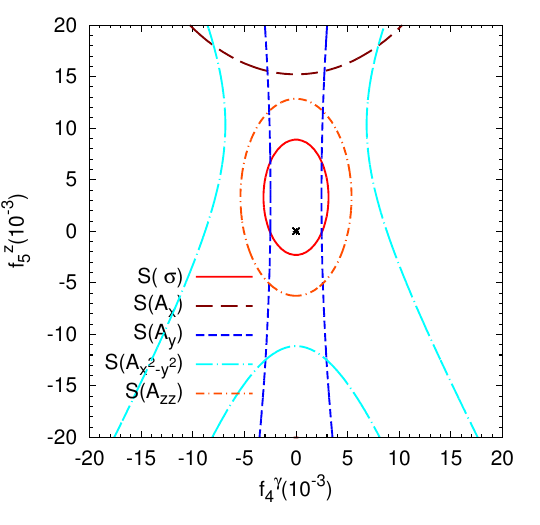}
\includegraphics[width=4.9cm]{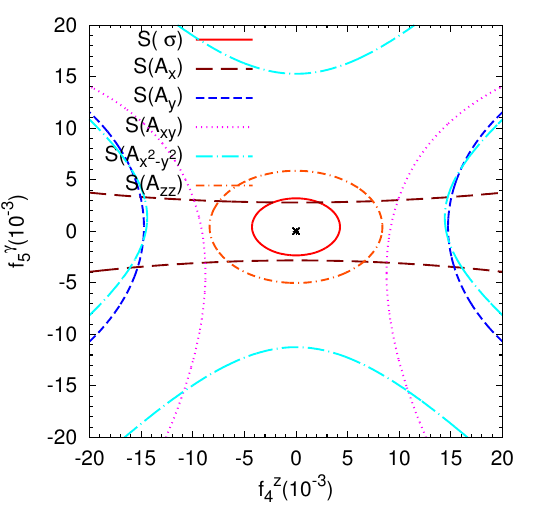}
\includegraphics[width=4.9cm]{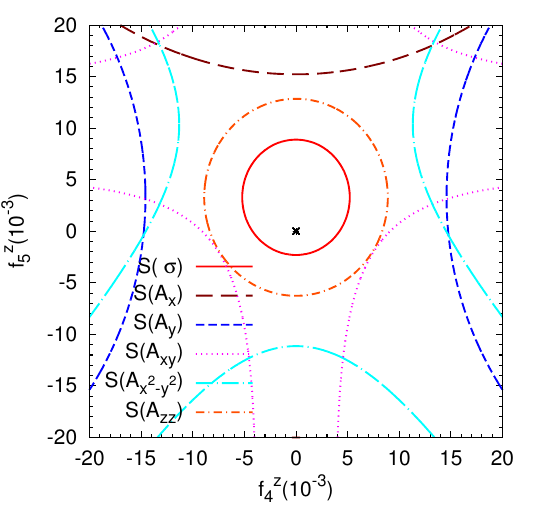}
\includegraphics[width=4.9cm]{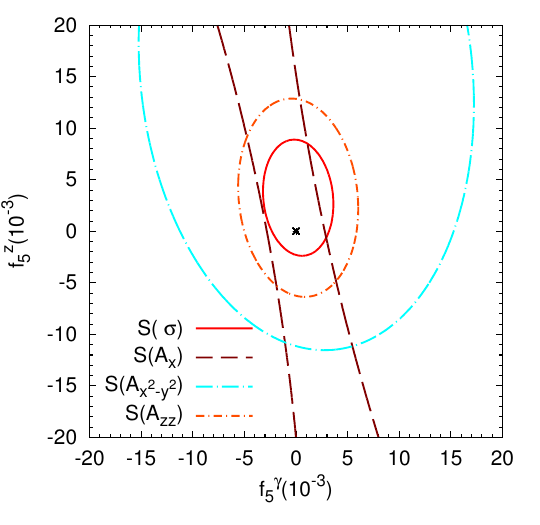}
\caption{\label{fig:Two_parameter_sensitivity_zz} $1\sigma$ sensitivity
contours ($\Delta\chi^2=1$) for  cross section and asymmetries obtained
by varying two parameters at a time and keeping the others at zero for the 
$ZZ$ process at $\sqrt{\hat{s}}=500$ GeV and $L=100$ 
fb$^{-1}$}
\end{figure*}

\begin{figure*}
\centering
\includegraphics[width=4.9cm]{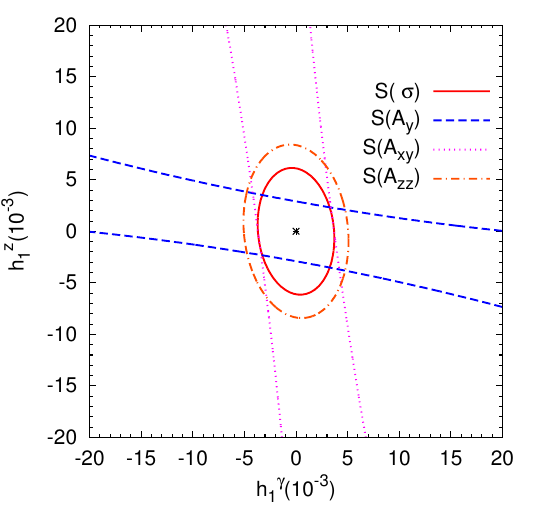}
\includegraphics[width=4.9cm]{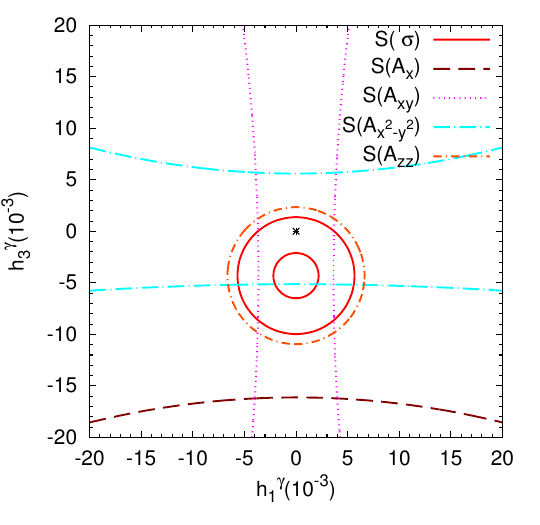}
\includegraphics[width=4.9cm]{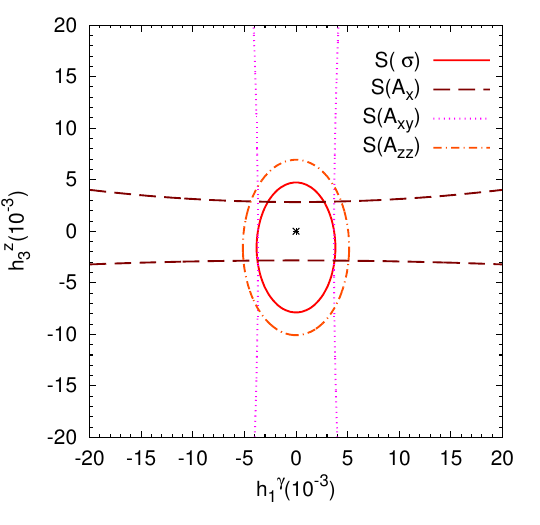}
\includegraphics[width=4.9cm]{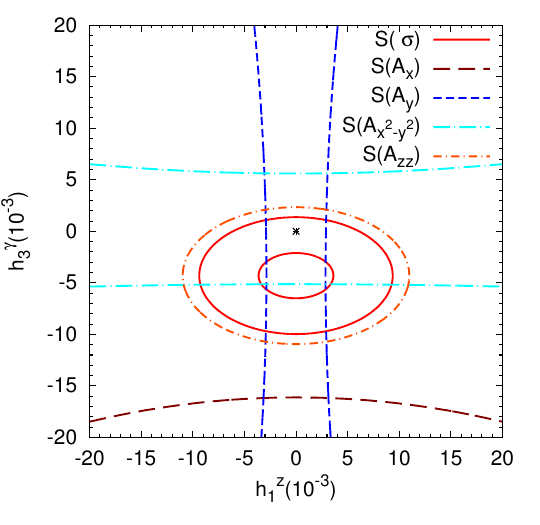}
\includegraphics[width=4.9cm]{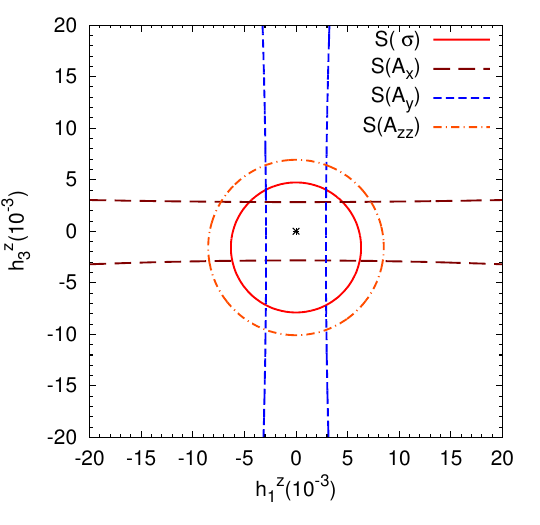}
\includegraphics[width=4.9cm]{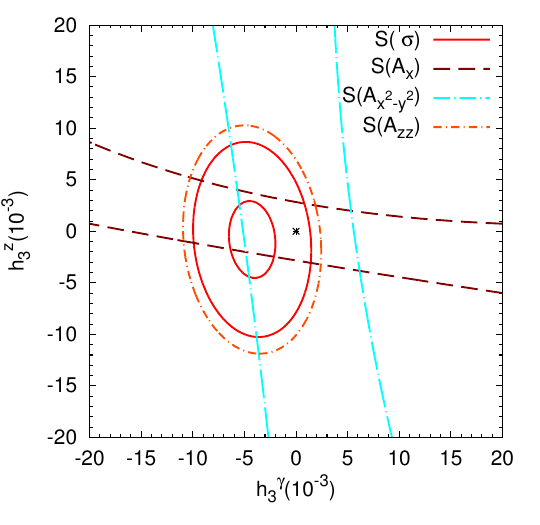}
\caption{\label{fig:Two_parameter_sensitivity_za}  $1\sigma$ sensitivity 
contours ($\Delta\chi^2=1$) for cross section and asymmetries obtained
by varying two parameters at a time and keeping the others at zero for the 
$Z\gamma$ process at $\sqrt{\hat{s}}=500$ GeV, $L=100$ fb$^{-1}$,
and $10^\circ \le\theta_\gamma\le170^\circ$}
\end{figure*}

We vary two couplings at a time, for each observable, and plot the 
${\cal S}=1$ (or $\Delta\chi^2=1$) contours in the corresponding parameter 
plane. These contours are shown in Fig.~\ref{fig:Two_parameter_sensitivity_zz}
and Fig.~\ref{fig:Two_parameter_sensitivity_za} for $ZZ$ and $Z\gamma$
processes, respectively. Asterisk ($\star$) marks in the middle of these plots
denote the SM value, i.e., the $(0,0)$ point. Each panel corresponds to two 
couplings
that are varied and all others are kept at zero. The shapes of the contours, for a
given observable, are a reflection of its dependence on the couplings as shown
in Tables~\ref{tab:parameter_dependent_pol_zz} and~\ref{tab:parameter_dependent_pol_za}. For example, let us look at the 
middle-top panel of Fig.~\ref{fig:Two_parameter_sensitivity_zz}, i.e. the
$(f_4^\gamma-f_5^\gamma)$ plane. The contours corresponding to the 
cross section (solid/red) and 
$A_{zz}$ (short-dash-dotted/orange) are circular in shape due to their quadratic
dependence on these two couplings with the same sign. The small linear dependence on
$f_5^\gamma$ makes these circles move toward a small positive value, as already
observed in the one-parameter analysis above. The $A_y$ contour 
(short-dash/blue) depends only on $f_4^\gamma$ in the numerator and a mild
dependence on $f_5^\gamma$ enters through the cross section, sitting in the
denominator of the asymmetries. The role of two couplings are exchanged for 
the $A_x$ contour (big-dash/black). The $A_{xy}$ contour (dotted/magenta) is 
hyperbolic in shape, indicating a dependence on the product 
$f_4^\gamma f_5^\gamma$, while a small shift toward positive $f_5^\gamma$ value
indicates a linear dependence on it. Similarly the symmetry about $f_4^\gamma=0$
indicates no linear dependence on it for $A_{xy}$. All these observations can
be confirmed by looking at  Table~\ref{tab:parameter_dependent_pol_zz} and
the expressions in \ref{appendix:Expresson_observables}. Finally, the shape of the $A_{x^2-y^2}$ contour 
(big-dash-dotted/cyan)
indicates a quadratic dependence on two couplings with opposite sign. Similarly,
all other panels can be read. Note that taking any one of the coupling
to zero in these panels gives us the $1\sigma$ limit on the other couplings as
found in the one-parameter analysis above.

In the contours for the $Z \gamma $ process,
Fig.~\ref{fig:Two_parameter_sensitivity_za}, one new kind of shape appears: the
annular ring corresponding to $\sigma_{Z\gamma}$ in middle-top, left-bottom, and right-bottom panels. This shape
corresponds to a large linear dependence of the cross section on $h_3^\gamma$
along with the quadratic dependence. By putting the other couplings to zero in
above-mentioned panels one obtains two disjoint internals for $h_3^\gamma$ at 
$1\sigma$ level as found before in the one-parameter analysis. The plane
containing two CP-odd couplings, i.e. the left-top panel, has two sets of
slanted contours corresponding to $A_{y}$ (short-dash/blue) and $A_{xy}$ 
(dotted/magenta), the CP-odd observables. These observables depend upon both
the couplings linearly and hence the slanted (almost) parallel lines. The rest of 
the panels can be read in the same way.

Till here we have used only one observable at a time for finding the limits.
A combination of all the observables would provide a much tighter limit on
the couplings than provided by any one of them alone. Also, the shape, the position, 
and the orientation of the allowed region would  change if the other two 
parameters were set to some value other than zero. A more comprehensive analysis
requires varying all the parameters and using all the observables to find the
parameter region of  low $\chi^2$ or high likelihood. The likelihood mapping of
the parameter space is performed using the MCMC method in
the next section.

\section{Likelihood mapping of parameter space}
\label{sec:MCMC}
In this section we perform a mock analysis of parameter estimation of anomalous coupling using {\em pseudo data} generated by {\tt MadGraph5}. We choose 
two benchmark points for coupling parameters as follows:
\begin{eqnarray}
\mbox{{\tt SM}} &:& f_{4,5}^V = 0.000, \ \ h_{1,3}^V = 0.000 \ \ 
\mbox{and}\nonumber\\
\mbox{{\tt aTGC}} &:& f_{4,5}^V = 0.005, \ \ h_{1,3}^V = 0.005 \ .
\end{eqnarray}
For each of these benchmark points we generate events in {\tt MadGraph5} 
 for {\em pseudo data} corresponding to ILC running at $500$ GeV and integrated 
 luminosity of $L=100$ fb$^{-1}$. The
likelihood of a given point $\vec{f}$ in the parameter space is defined by
\begin{equation}
{\cal L}(\vec{f}) = \prod_i \exp\left[- \ 
\frac{\left({\cal O}_i(\vec{f})-{\cal O}_i(\vec{f}_0)\right)^2}
{2\left(\delta {\cal O}_i(\vec{f}_0)\right)^2}\right]  ,
\end{equation}
where $\vec{f}_0$ defines the benchmark point. The product runs over the list
of observables we have: the cross section and five non-zero asymmetries.
We use the MCMC method to map the likelihood of the parameter space for each of the
benchmark point and for both processes. The one-dimensional marginalized 
distributions and the two-dimensional contours on the anomalous couplings are 
drawn from the Markov chains using the {\tt GetDist} package~\cite{Antony:GetDist}.

\subsection{MCMC analysis for $e^+e^-\to ZZ$}
Here we look at the process $e^+e^-\to ZZ$ followed by the decays $Z\to l^+l^-$
and $Z\to q\bar{q}$, with $l^-=e^-,~\mu^-$ in the \\ {\tt MadGraph5} simulations. 
The total cross section for this whole process would be
\begin{equation}
\sigma= \sigma(e^+e^-\to ZZ)\times 2 \ Br(Z\to l^+l^-) \ Br(Z\to q\bar{q}).
\end{equation}
The theoretical values of $\sigma(e^+e^-\to ZZ)$ and all the asymmetries are
obtained using expressions given in Appendix B and shown in the second column 
of Tables \ref{tab:marginalised_stat_observables_zz_sm} and 
\ref{tab:marginalised_observables_zz_bsm} for benchmark points {\tt SM} and
{\tt aTGC}, respectively. The {\tt MadGraph5} simulated values for these
observables are given in the third column of the two tables mentioned for
two benchmark points. Using these simulated values as {\em pseudo data} we
perform the likelihood mapping of the parameter space and obtain the posterior
distributions for parameters and the observables.
The last two columns of Tables \ref{tab:marginalised_stat_observables_zz_sm} and
\ref{tab:marginalised_observables_zz_bsm} show the 68~\% and 95~\% Bayesian
confidence interval (BCI) of the observables used. One naively expects  68~\%
BCI to roughly have the same size as the $1\sigma$ error in the {\em pseudo data}. 
However, we note that the  68~\% BCI for all the asymmetries is much narrower 
than expected, for both  benchmark points. This can be understood from the
fact that the cross section provides the strongest limit on any parameter, as
noticed in the earlier section, thus limiting the range of values for the
asymmetries. However, this must allow 68~\% BCI of the cross section to match with
the expectation. This indeed happens for the {\tt aTGC} case 
(Table~\ref{tab:marginalised_observables_zz_bsm}), but for the {\tt SM} case
even the cross section is narrowly constrained compared to a naive expectation.
The reason for this can be found in the dependence of the cross section on the
parameters. For most of the parameter space, the cross section is larger than
the {\tt SM} prediction and only for a small range of parameter space it can be
smaller. This was already pointed out while discussing multi-valued 
sensitivity in Fig.~\ref{fig:one_parameter_sensitivity_zz}. We found the
lowest possible value of the cross section to be $37.77$ fb, obtained for 
$f_4^{\gamma,Z}\approx 0$, $f_5^\gamma \sim 2\times10^{-4}$, and $f_5^Z \sim
3.2\times10^{-3}$.
\begin{table*}
\centering
\caption{\label{tab:marginalised_stat_observables_zz_sm} 
List of observables shown for the process $e^+e^-\to ZZ$ for the  benchmark 
point {\tt SM} with $\sqrt{\hat{s}}=500$ GeV: theoretical values (column 2), 
{\tt MadGraph5} simulated value for $L=100$ fb$^{-1}$ (column 3), 68~\% (column
4) and 95~\% (column 5) Bayesian confidence intervals (BCI) }
\renewcommand{\arraystretch}{1.5}
\begin{tabular*}{\textwidth}{@{\extracolsep{\fill}}lllll@{}}\hline
Observables    &Theoretical ({\tt SM}) & MadGraph ({\tt SM}, prior)  
&68~\% BCI (posterior)& 95~\% BCI (posterior)\\ 
\hline
$\sigma$      & $ 38.096$ fb  & $ 38.16\pm 0.62 $ fb    &
$38.61^{+0.31}_{-0.53}$fb        & $38.61^{0.83}_{-0.74}$fb     \\ 
$A_x$         & $ 0.00099$ & $0.0023\pm 0.0161$   & $-0.0021\pm 0.0087
$     & $-0.0021^{+0.016}_{-0.017}  $    \\
$A_y$         & $0 $           & $-0.0016\pm 0.0161 $ & $-0.0005\pm 0.0090
$     & $-0.0005^{+0.017}_{-0.017}  $    \\ 
$A_{xy}$      & $0 $           & $0.0004\pm 0.0161$  & $0.0001\pm 0.0036
$     & $0.0001^{+0.0071}_{-0.0071}$      \\ 
$A_{x^2-y^2}$ & $-0.02005 $  & $-0.0189\pm 0.0161$   &
$-0.0166^{+0.0032}_{-0.0018}$    & $-0.0166^{+0.0043}_{-0.0052}$     \\ 
$A_{zz}$      & $0.17262 $    & $0.1745\pm 0.0159 $    &
$0.1691^{+0.0035}_{-0.0022}$     & $0.1691^{+0.0051}_{-0.0056}$      \\
\hline
\end{tabular*}
\end{table*}
Thus, for  most of the parameter space the anomalous couplings cannot 
emulate the negative statistical fluctuations in the cross section making the
likelihood function, effectively, a one-sided Gaussian function. This 
forces the mean of posterior distribution to a higher value. We also note that 
the upper bound of the 68~\% BCI for cross section ($38.92$ fb) is comparable to 
the expected $1\sigma$ upper bound ($38.78$ fb). Thus we have an overall 
narrowing of the range of the posterior
distribution of the cross section values. This, in turns, leads to a narrow range
of parameters allowed and hence narrow ranges for the asymmetries in the case 
of {\tt SM} benchmark point. For the {\tt aTGC} benchmark point, it is possible
to emulate the negative fluctuations in cross section by varying the 
parameters, thus the corresponding posterior distributions compare with the 
expected $1\sigma$ fluctuations. The narrow ranges for the posterior distribution 
for all the asymmetries are due to the tighter constraints on the parameters coming
from the cross section and correlation between the observables. 

\begin{table*}
\centering
\caption{\label{tab:marginalised_observables_zz_bsm} 
List of observables shown for the process $e^+e^-\to ZZ$ for the  benchmark 
point {\tt aTGC} with $\sqrt{\hat{s}}=500$ GeV. The rest of the details are the same as in 
Table~\ref{tab:marginalised_stat_observables_zz_sm} }
\renewcommand{\arraystretch}{1.5}
\begin{tabular*}{\textwidth}{@{\extracolsep{\fill}}lllll@{}}\hline
Observables    &Theoretical ({\tt aTGC})& MadGraph ({\tt aTGC}, prior) & 
68~\% BCI (posterior) 
& 95~\% BCI (posterior) \\ \hline
$\sigma$      & $43.307 $ fb & $43.33\pm 0.6582 $ fb      & $43.40\pm 0.66
$ fb        & $43.40 \pm 1.3$ fb         \\  
$A_x$         & $-0.02954 $ & $-0.0308\pm 0.0151$    &
$-0.0240^{+0.0087}_{-0.013}$ & $-0.0240^{+0.021}_{-0.020}  $   \\ 
$A_y$         & $0.03424 $  & $0.0308\pm 0.0151$     &
$0.0230^{+0.013}_{-0.0085} $ & $0.0230^{+0.020}_{-0.022}   $    \\  
$A_{xy}$      & $ 0.00574$ & $0.0056\pm 0.0152$    &
$0.0041^{+0.0076}_{-0.0063}$ & $0.0041^{+0.015}_{-0.015}   $    \\
$A_{x^2-y^2}$ & $-0.00941 $ & $-0.0119 \pm 0.0152$   &
$-0.0116^{+0.0071}_{-0.0032}$& $-0.0116^{+0.0093}_{-0.012}$     \\
$A_{zz}$      & $ 0.14057$   & $0.1382 \pm 0.0150$     & $0.1401\pm 0.0035
$ & $0.1401^{+0.0069}_{-0.0067}$     \\\hline
\end{tabular*}
\end{table*}
\begin{figure*}
\centering
\includegraphics[width=3.6cm]{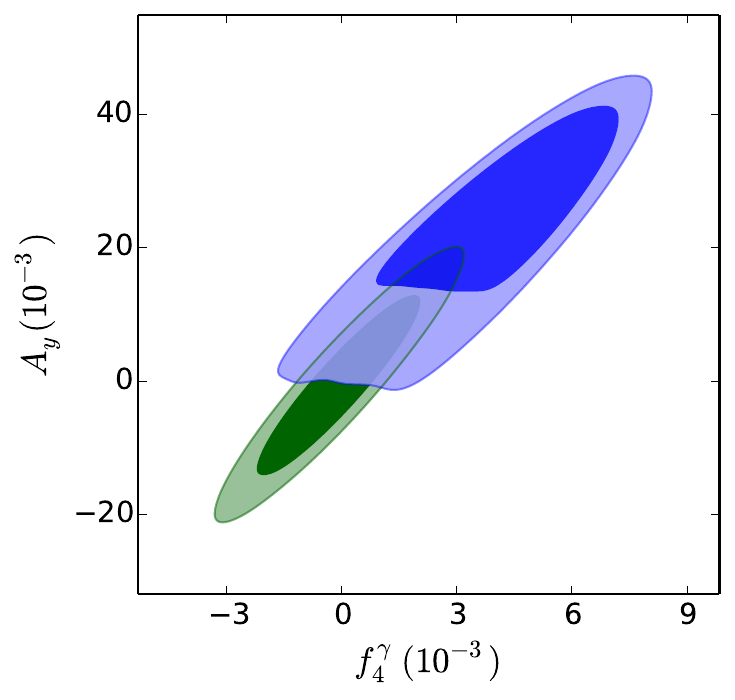}
\includegraphics[width=3.6cm]{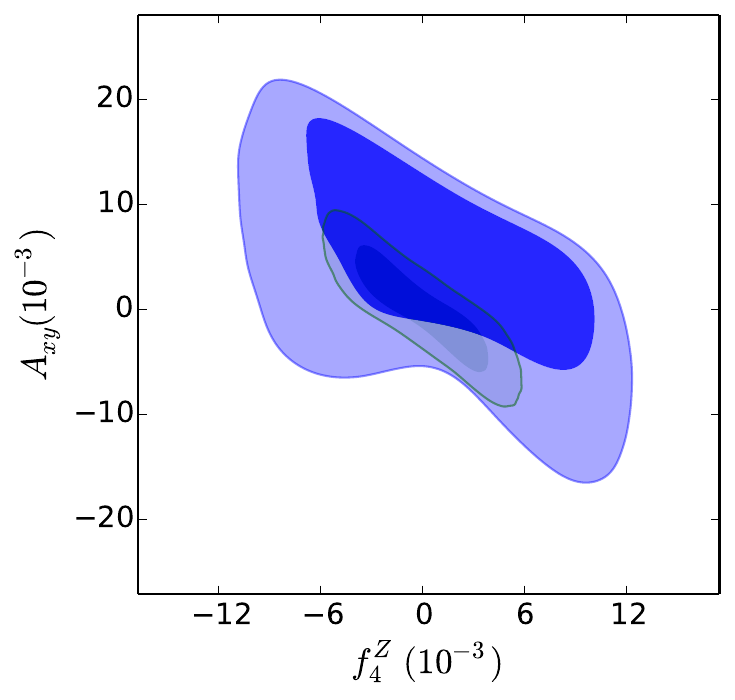}
\includegraphics[width=3.6cm]{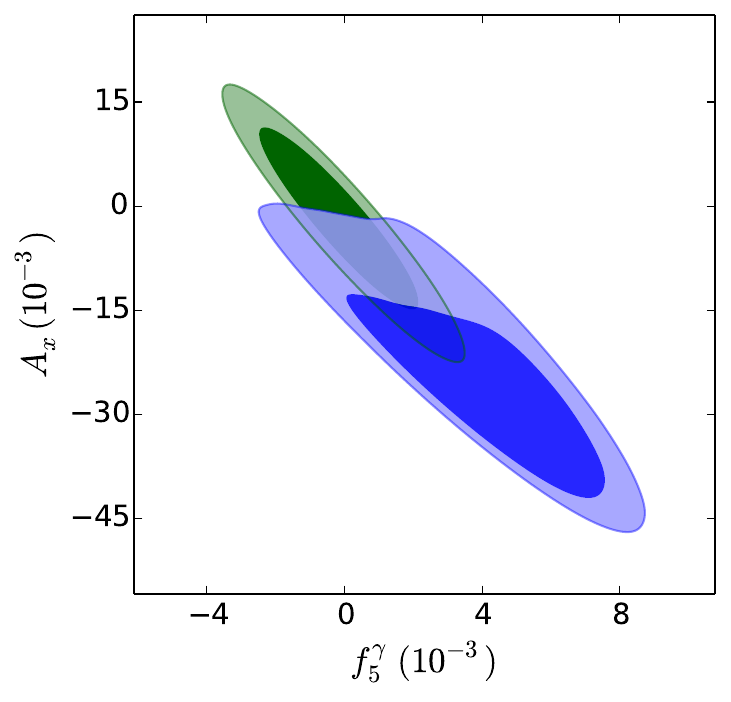}
\includegraphics[width=3.6cm]{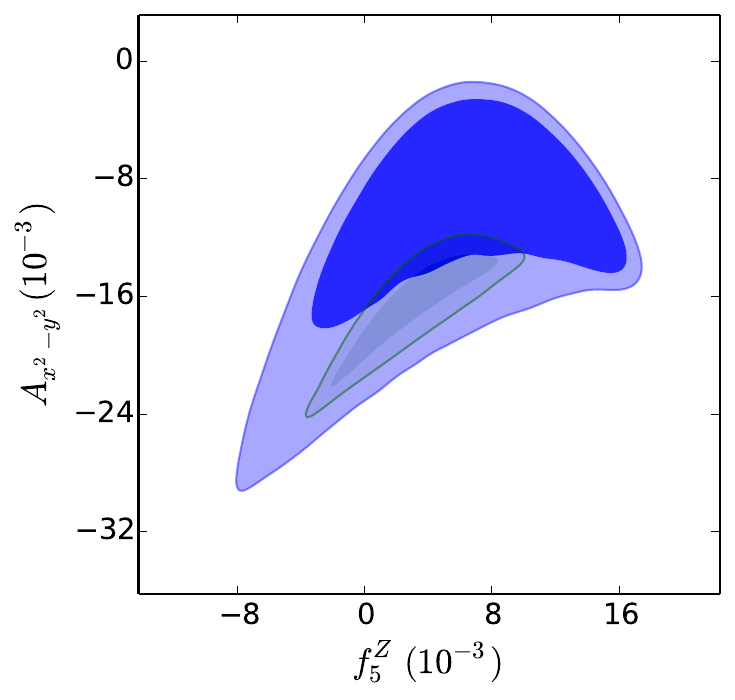}
\caption{\label{fig:correlations_observables_in_mcmc_zz} 
Two-dimensional marginalized contours showing most correlated observable for
each parameter of the  process $e^+e^-\rightarrow ZZ$. The upper transparent layer ({\em blue})
contours correspond to {\tt aTGC}, while the lower layer ({\em green}) contours
correspond to {\tt SM}. The darker shade shows 68~\% contours, while the lighter
shade is for 95~\% contours}
\end{figure*}
\begin{figure}
\centering
\includegraphics[width=3.6cm]{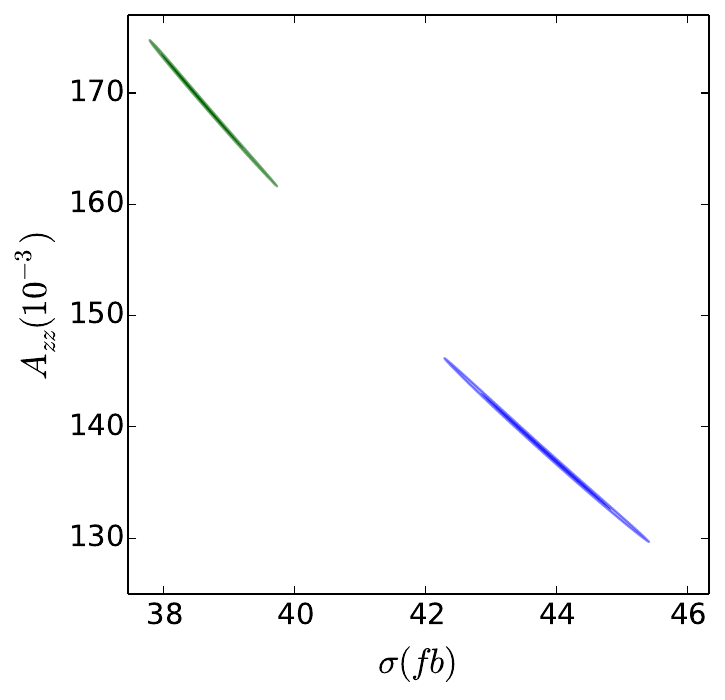}
\caption{\label{fig:sigmatzz_zz_withc} Two-dimensional marginalized contours showing correlation between $A_{zz}$ and $\sigma$ in the $ZZ$ process. The rest of the details are the same as in Fig.~\ref{fig:correlations_observables_in_mcmc_zz}}
\end{figure}

We are using a total of six observables, five asymmetries and one cross section
for our analysis of two benchmark points; however, we have only four free 
parameters. This invariably leads to some correlations between the observables
apart from the expected correlations between parameters and observables.
Figure~\ref{fig:correlations_observables_in_mcmc_zz} shows
most prominently correlated observable for each of the parameters. The CP
nature of observables is reflected in the parameter it is strongly correlated
with. We see that $A_y$ and $A_{xy}$ are linearly dependent upon both $f_4^\gamma$ and
$f_4^Z$; however, $A_y$ is more sensitive to $f_4^\gamma$ as shown in 
Fig.~\ref{fig:one_parameter_sensitivity_zz} as well. Similarly, for the other
asymmetries and parameters one can see a correlation which is consistent with the
sensitivity plots in Fig.~\ref{fig:one_parameter_sensitivity_zz}.
The strong (and negative) correlation between $A_{zz}$ and $\sigma$ shown in Fig.~\ref{fig:sigmatzz_zz_withc} 
indicates that any one of them is sufficient for the analysis, in principle.
However, in practice the cross section puts a much stronger limit than $A_{zz}$, which
explains the much narrower BCI for it as compared to the $1\sigma$ expectation.
\begin{table*}
\centering
\caption{\label{tab:marginalised_parameters_zz} 
The list of best-fit points, posterior 68~\% and 95~\% BCI  for the parameters for 
the process $e^+e^-\to ZZ$ for both {\tt SM} and {\tt aTGC} benchmark points
}
\renewcommand{\arraystretch}{1.5}
\begin{tabular*}{\textwidth}{@{\extracolsep{\fill}}lllllll@{}}\hline
\multicolumn{1}{c}{}&\multicolumn{3}{c}{{\tt SM} Benchmark}
& \multicolumn{3}{c}{{\tt aTGC} Benchmark}\\ \hline
 $f^V_i$ &   68~\% BCI &  95~\% BCI & Best-fit & 68~\% BCI  &  95~\% BCI & Best-fit
\\\hline
$f_4^\gamma$   & $-0.0001\pm 0.0014  $  & $-0.0001^{+0.0027}_{-0.0027}$
&$-0.0002$ & $0.0038^{+0.0026}_{-0.0016}$& $0.0038^{+0.0037}_{-0.0042}$&$0.0044$
\\ 
$f_4^Z$        & $0.0000\pm 0.0026   $  & $0.0000^{+0.0049}_{-0.0049}$
&$-0.0002$  & $0.0010^{+0.0065}_{-0.0055}$& $0.0010^{+0.0098}_{-0.011} $
&$0.0050$  \\ 
$f_5^\gamma$   & $-0.0001\pm 0.0015  $  & $-0.0001^{+0.0030}_{-0.0029}$ &
$-0.0002$ & $0.0038^{+0.0029}_{-0.0019}$& $0.0038^{+0.0042}_{-0.0047}$ &$0.0057$
\\ 
$f_5^Z$        & $0.0032\pm 0.0028   $  & $0.0032^{+0.0053}_{-0.0053}$
&$0.0000$  & $0.0057^{+0.0074}_{-0.0051}$& $0.0057^{+0.010}_{-0.011}
$&$0.0037$  \\ \hline
\end{tabular*}
\end{table*}
\begin{figure*}
\centering
\includegraphics[width=3.6cm]{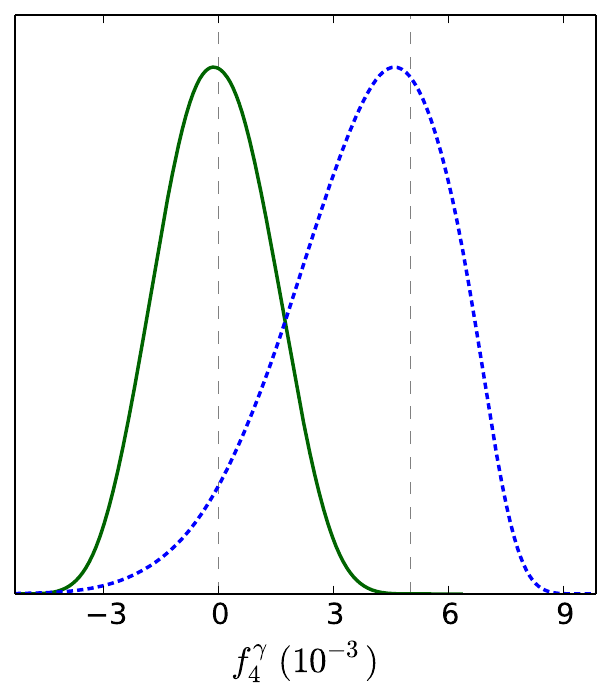}
\includegraphics[width=3.6cm]{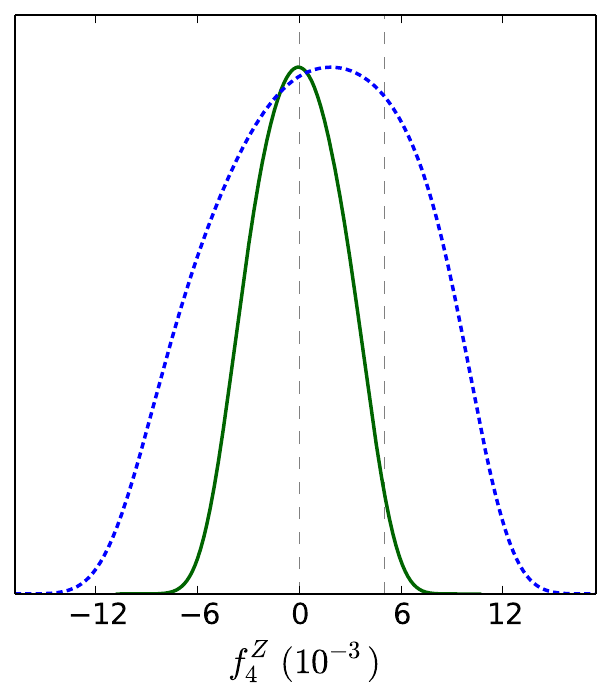}
\includegraphics[width=3.6cm]{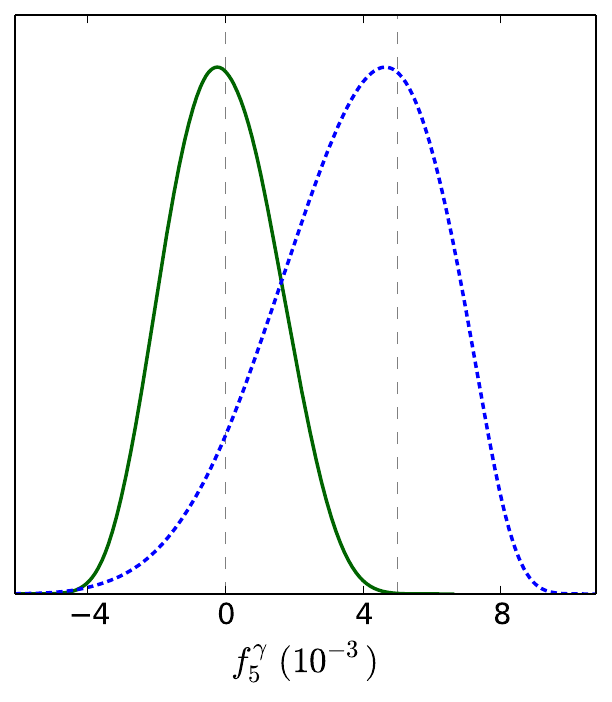}
\includegraphics[width=3.6cm]{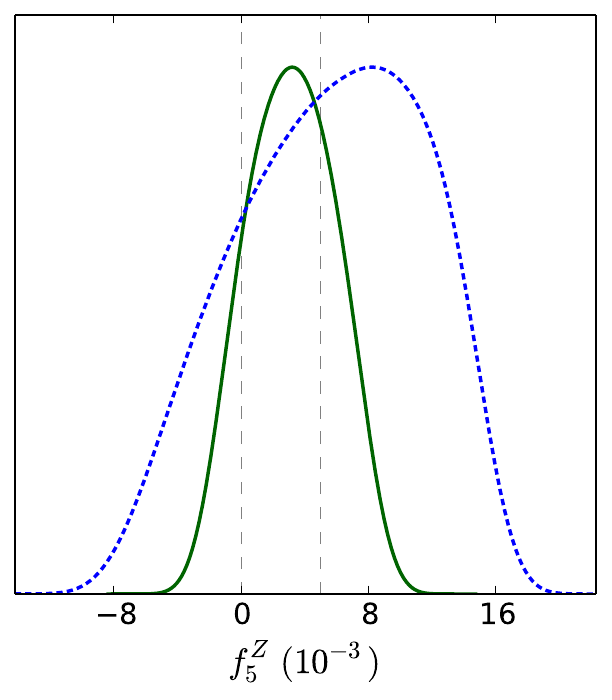}
\caption{\label{fig:one_parameter_mcmc_parameter_zz} 
One-dimensional marginalized posterior distribution for the parameters of the 
process $e^+e^-\rightarrow ZZ$. {\em Solid (green)} lines are for {\tt SM} and 
{\em dashed (blue)} lines are for {\tt aTGC} hypothesis. The values of the parameters
for the benchmark points are shown by vertical lines for reference}
\end{figure*}
\begin{figure*}
\centering
\includegraphics[width=4.9cm]{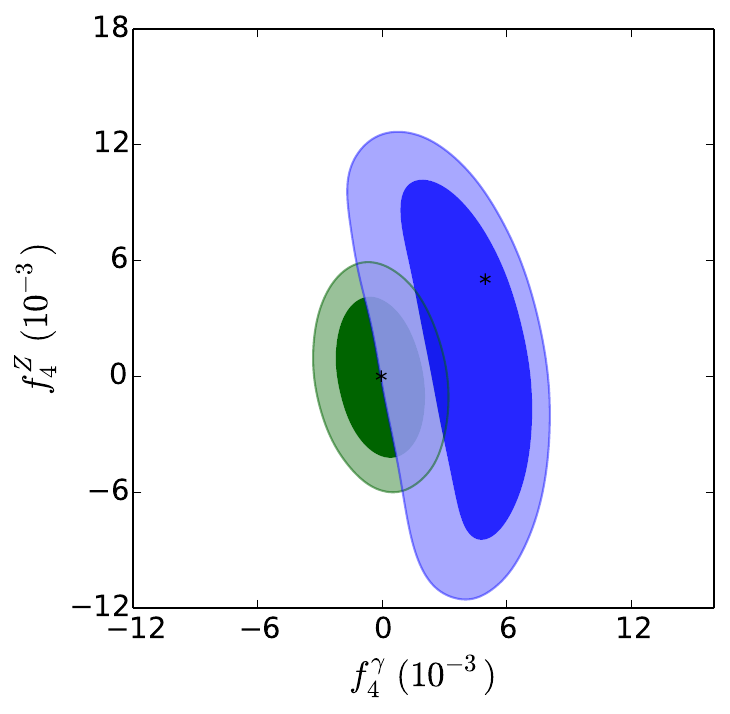}
\includegraphics[width=4.9cm]{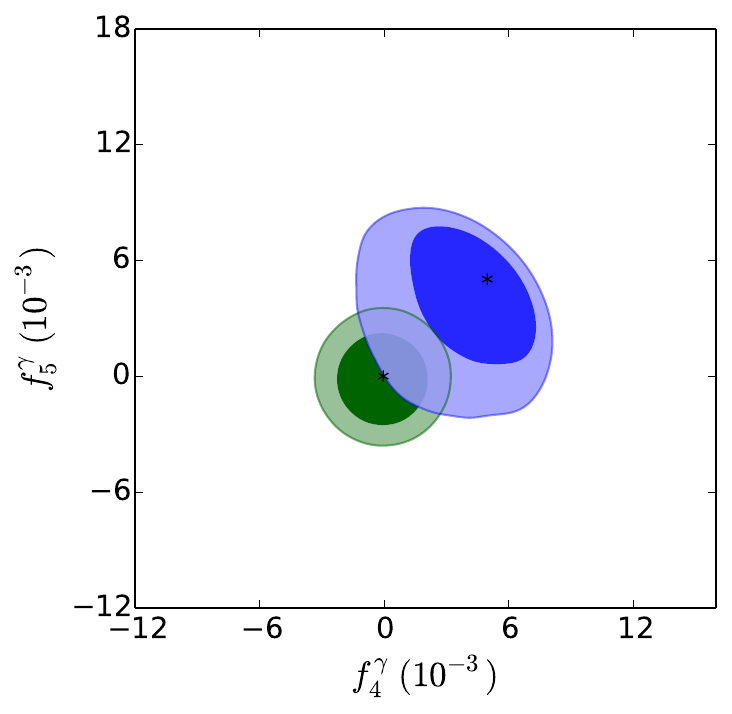}
\includegraphics[width=4.9cm]{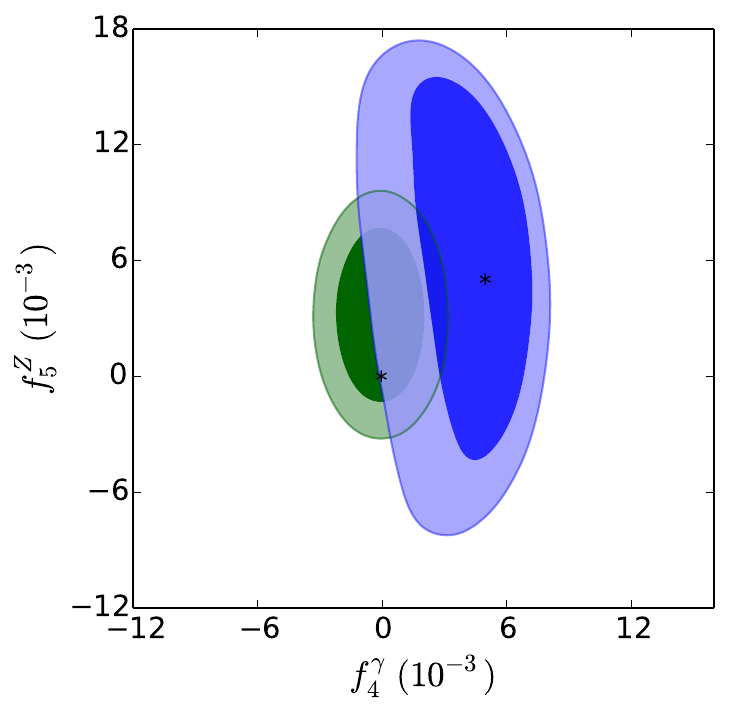}
\includegraphics[width=4.9cm]{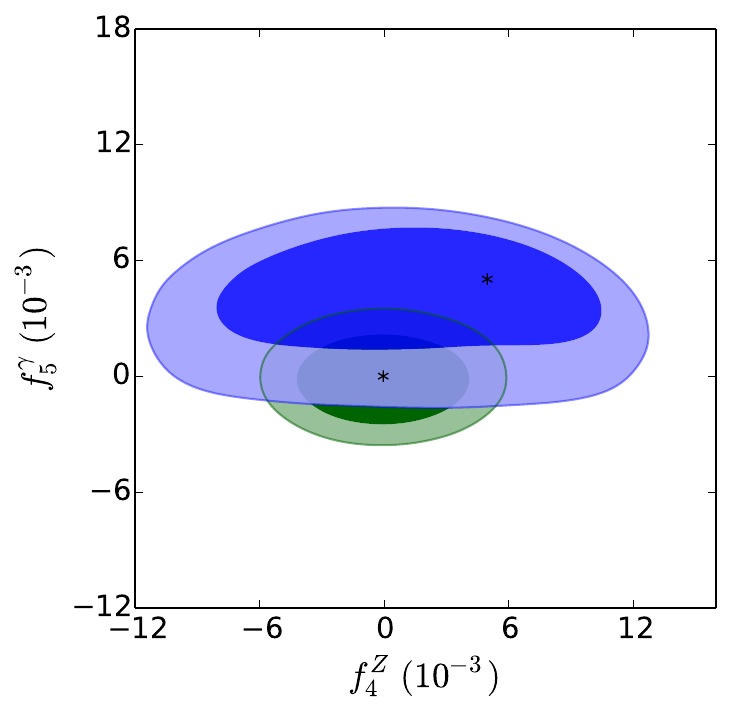}
\includegraphics[width=4.9cm]{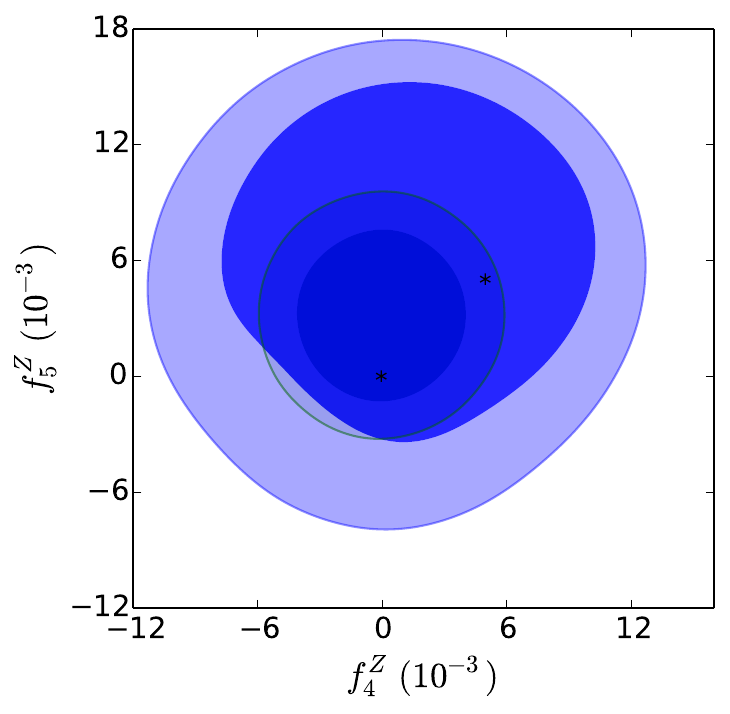}
\includegraphics[width=4.9cm]{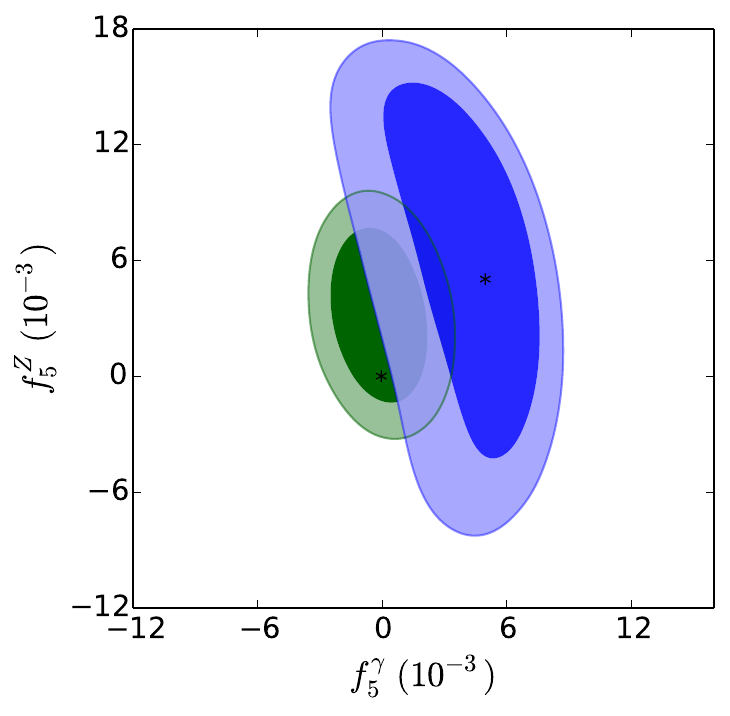}
\caption{\label{fig:correlations_parameters_in_mcmc_zz} 
Two-dimensional marginalized contours showing correlations between parameters
of the  process $e^+e^-\rightarrow ZZ$. The other details are the same as in 
Fig.~\ref{fig:correlations_observables_in_mcmc_zz}.
}
\end{figure*}

Finally, we come to the discussion of the parameter  estimation. The marginalized
one-dimensional posterior distributions for the parameters of $ZZ$ production
process are shown in Fig.~\ref{fig:one_parameter_mcmc_parameter_zz}, while
the corresponding BCI along with best-fit points are listed in 
Table~\ref{tab:marginalised_parameters_zz} for both  benchmark points. 
The vertical lines near zero correspond to the true value of parameters for 
{\tt SM} and the other vertical lines correspond to {\tt aTGC}. The best-fit
points are very close to the true values except for $f_5^Z$ in the  {\tt aTGC} 
benchmark point due
to the multi-valuedness of the cross section. The 95~\% BCI of the parameters for two 
benchmark points overlap and it appears as if they cannot be resolved. To see 
the resolution better we plot two-dimensional posteriors in 
Fig.~\ref{fig:correlations_parameters_in_mcmc_zz}, with the benchmark points
shown with an asterisk. Again we see that the 95~\% contours do overlap as these
contours are obtained after marginalizing over non-shown parameters in each 
panel. Any higher-dimensional representation is not possible on paper, but we
have checked three-dimensional scatter plot of points on the Markov chains and 
conclude that the shape of the {\em good likelihood} region is ellipsoidal for
the {\tt SM} point with the true value at its centre. The corresponding
three-dimensional
shape for the {\tt aTGC} point is like a part of an ellipsoidal shell. Thus
in full four-dimension there will not be any overlap 
(see Section \ref{ssec:separation})
and we can distinguish the
two chosen benchmark points as it is quite obvious from the corresponding 
cross sections. However, left to only the cross section we would have the 
entire ellipsoidal shell as possible range of parameters for the {\tt aTGC} 
case. The presence of asymmetries in our analysis helps narrow down to a part
of the ellipsoid and hence aids the parameter estimation for the $ZZ$ production
process.

\subsection{MCMC analysis for $e^+e^-\to Z\gamma$ }
Next we look at the process $e^+e^-\to Z\gamma$ and $Z\to l^+l^-$ with 
$l^-=e^-, \mu^-$ in the {\tt MadGraph5} simulations. The total cross section 
for this whole process is given by
\begin{equation}
\sigma= \sigma(e^-e^+\to Z\gamma)\times Br(Z\to l^+l^-).
\end{equation}
The theoretical values of the cross section and asymmetries (using expression 
in~\ref{appendix:Expresson_observables}) are given in the second column of
Tables \ref{tab:marginalised_observables_za_sm} and
\ref{tab:marginalised_observables_za_bsm} for {\tt SM} and {\tt aTGC} points,
respectively. The tables contain the {\tt MadGraph5} simulated data for 
$L=100$ fb$^{-1}$ along with 68~\% and 95~\% BCI for the observables obtained from
the MCMC analysis. For the {\tt SM} point, 
Table~\ref{tab:marginalised_observables_za_sm}, we notice that the 68~\% BCI for
all the observables are narrower than the $1\sigma$ range of the {\em psuedo
data} from {\tt MadGraph5}. This is again related to the correlations between
observables and the fact that the cross section has a lower bound of about 111
fb obtained for $h_3^\gamma\sim -4.2\times 10^{-3}$ with other parameters close
to zero. This lower bound of the cross section leads to narrowing of 68~\% BCI for
$\sigma$ and hence for other asymmetries too, as observed in the $ZZ$ 
production process. The 68~\% BCI for 
$A_{x^2-y^2}$ and $A_{zz}$ are particularly narrow. For $A_{zz}$, this is 
related to the strong correlation between $(\sigma - A_{zz})$, while for 
$A_{x^2-y^2}$ the slower dependence on $h_3^\gamma$ along with strong 
dependence of $\sigma$ on $h_3^\gamma$ is the cause of a narrow 68~\% BCI.
\begin{table*}
\centering
\caption{\label{tab:marginalised_observables_za_sm}
List of observables shown for the process $e^+e^-\to Z \gamma$ for the {\tt SM} 
point with $\sqrt{\hat{s}}=500$ GeV and $L=100$ fb$^{-1}$. The rest of the details are the 
same as in Table~\ref{tab:marginalised_stat_observables_zz_sm}
}
\renewcommand{\arraystretch}{1.5}
\begin{tabular*}{\textwidth}{@{\extracolsep{\fill}}lllll@{}}\hline
 Observables    & Theoretical ({\tt SM})& MadGraph ({\tt SM}, prior) & 
68~\% BCI (posterior) & 95~\% BCI (posterior)  \\ \hline
$\sigma$      & $ 112.40$ fb & $112.6\pm 1.06 $  fb   &
$112.64^{+0.64}_{-0.91}$ fb      & $112.6^{+1.5}_{-1.4}$ pb\\
$A_x$         & $0.00480 $ & $0.0043\pm 0.0094 $ & $0.0041\pm 0.0088
$     & $0.0041^{+0.017}_{-0.018}   $ \\
$A_y$         & $0 $          & $-0.0011\pm 0.0094$ & $-0.0009\pm 0.0088
$     & $-0.0009 \pm 0.017  $ \\
$A_{xy}$      & $ 0$          & $0.0003\pm 0.0094$ & $0.0001\pm 0.0065
$     & $0.0001 \pm 0.012  $ \\
$A_{x^2-y^2}$ & $0.00527 $ & $0.0056\pm 0.0094$ &
$-0.0001^{+0.0064}_{-0.0034}$    & $-0.0001^{+0.0079}_{-0.0096}$\\
$A_{zz}$      & $ 0.17819$   & $0.1781\pm 0.0092$   &
$0.1771^{+0.0043}_{-0.0031}$     & $0.1771^{+0.0066}_{-0.0070}$ \\\hline
\end{tabular*}
\end{table*}
\begin{table*}
\centering
\caption{\label{tab:marginalised_observables_za_bsm} 
List of observables shown for the process $e^+e^-\to Z\gamma$ for the 
{\tt aTGC} point with $\sqrt{\hat{s}}=500$ GeV and $L=100$ fb$^{-1}$. The rest of the 
details are the same as in Table~\ref{tab:marginalised_stat_observables_zz_sm}
}
\renewcommand{\arraystretch}{1.5}
\begin{tabular*}{\textwidth}{@{\extracolsep{\fill}}lllll@{}}\hline
Observables & Theoretical ({\tt aTGC})&MadGraph ({\tt aTGC}, prior) & 
68~\% BCI (posterior) & 95~\% BCI  (posterior)\\\hline
$\sigma$      & $122.0 $ fb  & $122.4 \pm 1.11$ fb   &  $122.3\pm 1.0
$ fb    & $122.3 \pm 2.0$ fb          \\
$A_x$         & $0.02404 $   & $0.0252\pm 0.0090$  &  $0.0263\pm 0.0093
$   & $0.0263 \pm 0.018   $        \\
$A_y$         & $-0.01775 $  & $-0.0165\pm 0.0090$ &  $-0.0172\pm 0.0092
$   & $-0.0172 \pm 0.018  $        \\
$A_{xy}$      & $ -0.01350$  & $-0.0104\pm 0.0090$ &
$-0.0109^{+0.0069}_{-0.011}$   & $-0.0109^{+0.017}_{-0.015}  $\\
$A_{x^2-y^2}$ & $0.01440 $   & $0.0133\pm 0.0090$  &
$0.0121^{+0.0055}_{-0.0010}$   & $0.0121^{+0.0068}_{-0.012} $\\
$A_{zz}$      & $0.13612 $    & $0.1361\pm 0.0089$   &  $0.1351\pm 0.0041
$   & $0.1351^{+0.0080}_{-0.0079}$\\\hline
\end{tabular*}
\end{table*}

\begin{figure*}
\centering
\includegraphics[width=3.6cm]{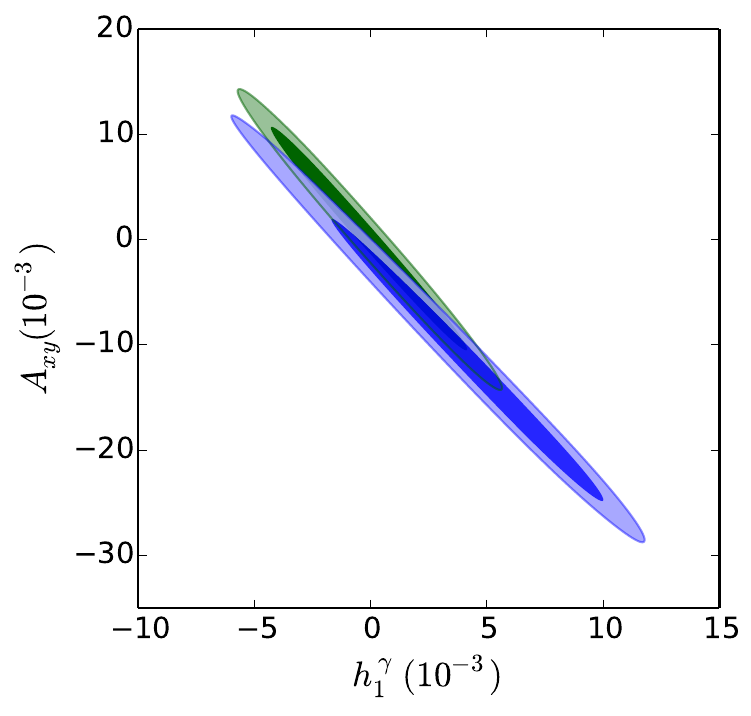}
\includegraphics[width=3.6cm]{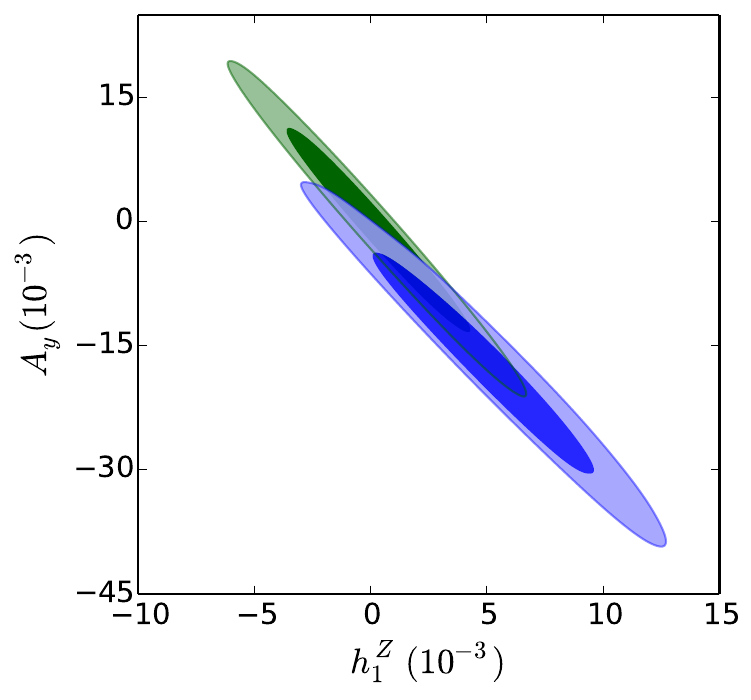}
\includegraphics[width=3.6cm]{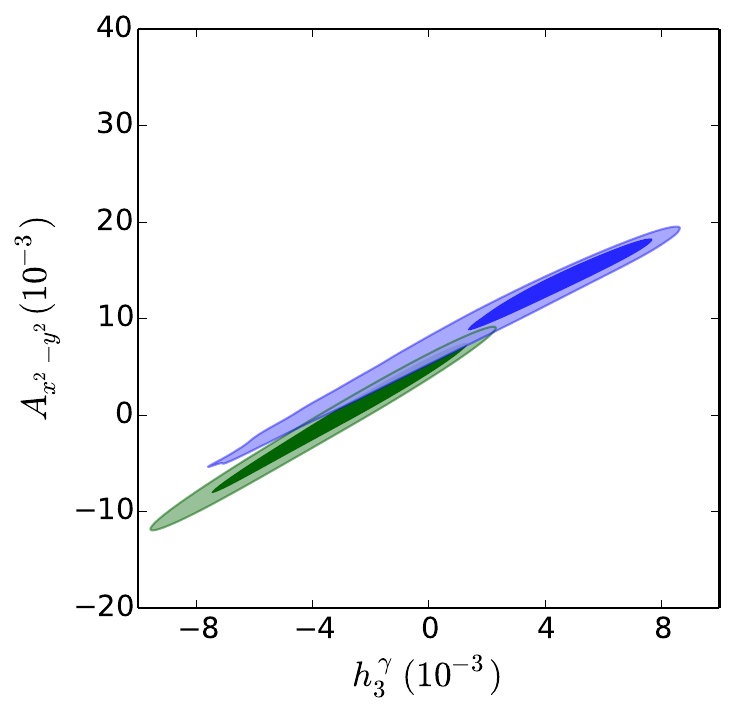}
\includegraphics[width=3.6cm]{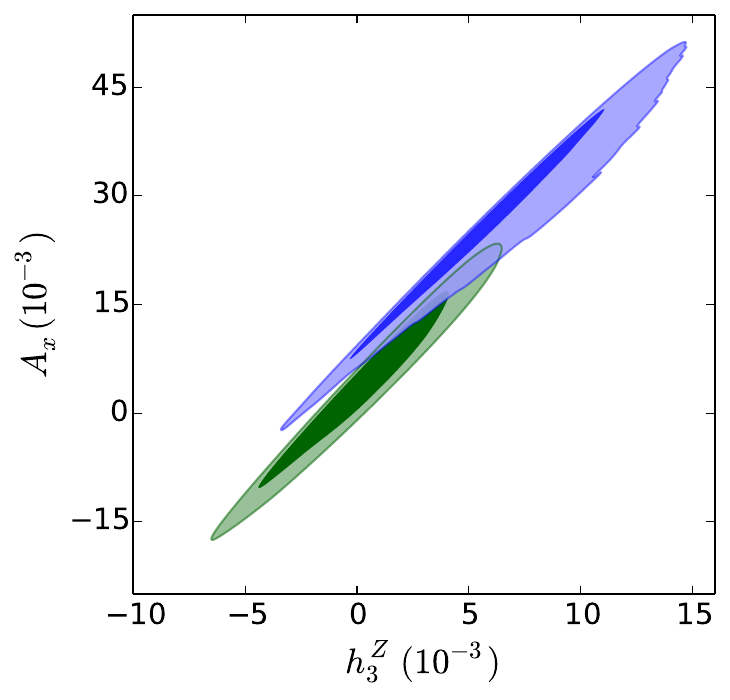}
\caption{\label{fig:correlations_observables_in_mcmc_za} 
Two-dimensional marginalized contours showing most correlated observables for each
parameter of the process $e^+e^-\rightarrow Z\gamma$ for two benchmark points. 
The rest of the details are the same as in
Fig.~\ref{fig:correlations_observables_in_mcmc_zz}
}
\end{figure*}

For the {\tt aTGC} point, there is enough room for the negative fluctuation in
the cross section and hence  no narrowing of the 68~\% BCI is observed
for it; see Table~\ref{tab:marginalised_observables_za_bsm}. The 68~\% BCI for
$A_x$ and $A_y$ are comparable to the corresponding $1\sigma$ intervals, while
the  68~\% BCI for other three asymmetries are certainly narrower than 
$1\sigma$ intervals. This narrowing, as discussed earlier, is due to the 
parametric dependence of the observables and their correlations.  
Each of the parameters has a strong correlation with one of the asymmetries as
shown in Fig.~\ref{fig:correlations_observables_in_mcmc_za}. The narrow 
contours indicate that if one can improve the errors on the asymmetries, it will 
improve the parameter extraction. The steeper is the slope of the 
narrow contour, the larger will be its improvement. We note that $A_x$ and $A_y$
 have a steep dependence on the corresponding parameters, thus even small 
variations in the parameters lead to large variations in the asymmetries. 
For $A_{xy}$ and $A_{x^2-y^2}$ the parametric dependence is weaker, leading to
their smaller variation with the parameters and hence narrower 68~\% BCI.

\begin{table*}
\centering
\caption{\label{tab:marginalised_parameters_za_sm} 
The list of best-fit points, posterior 68~\% and 95~\% BCI  for the parameters for the process
$e^+e^-\to Z\gamma$ for both  benchmark points
}
\renewcommand{\arraystretch}{1.5}
\begin{tabular*}{\textwidth}{@{\extracolsep{\fill}}lllllll@{}}\hline
\multicolumn{1}{c}{}&\multicolumn{3}{c}{{\tt SM} Benchmark}
& \multicolumn{3}{c}{{\tt aTGC} Benchmark}\\ \hline
 $h^V_i$    &  68~\% BCI                     &  95~\% BCI &Best-fit
&  68~\% BCI &  95~\% BCI & Best-fit  \\ \hline
$h_1^\gamma$  & $-0.0001\pm 0.0026         $     & $-0.0001^{+0.0048}_{-0.0047}$
&$ -0.0002$ & $0.0039^{+0.0047}_{-0.0031}$   & $0.0039^{+0.0068}_{-0.0075}$
&$0.0040 $\\
$h_1^Z$       & $0.0003\pm 0.0028          $     & $0.0003^{+0.0054}_{-0.0054}$
& $0.0001 $& $0.0050\pm 0.0033          $   &
$0.0050^{+0.0064}_{-0.0063}$&$0.0047 $\\
$h_3^\gamma$  & $-0.0030^{+0.0036}_{-0.0020}$    & $-0.0030^{+0.0045}_{-0.0054}$
& $0.0002 $& $0.00348^{+0.0036}_{-0.00086}$ &
$0.00348^{+0.0047}_{-0.0076}$&$0.0056 $\\
$h_3^Z$       & $0.0004\pm 0.0028          $     & $0.0004^{+0.0053}_{-0.0055}$
& $-0.0002 $& $0.0062^{+0.0030}_{-0.0035}$   &
$0.0062^{+0.0070}_{-0.0062}$&$0.0052 $\\\hline
\end{tabular*}
\end{table*}
\begin{figure*}
\centering
\includegraphics[width=3.6cm]{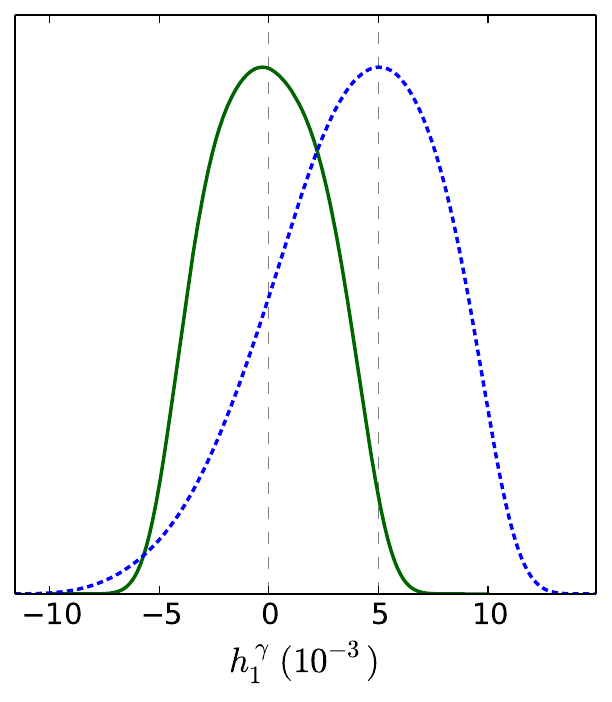}
\includegraphics[width=3.6cm]{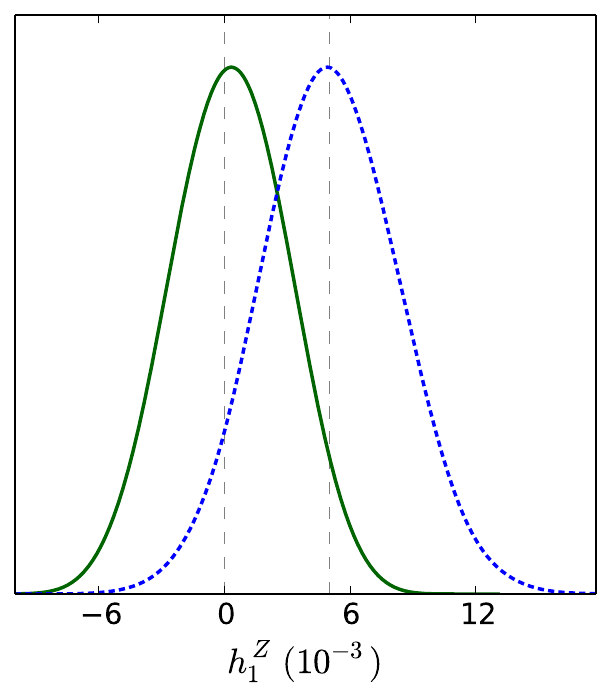}
\includegraphics[width=3.6cm]{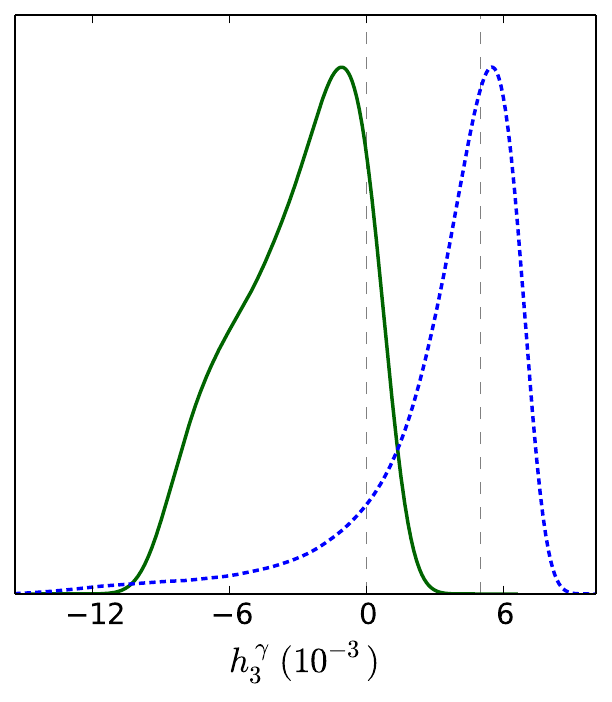}
\includegraphics[width=3.6cm]{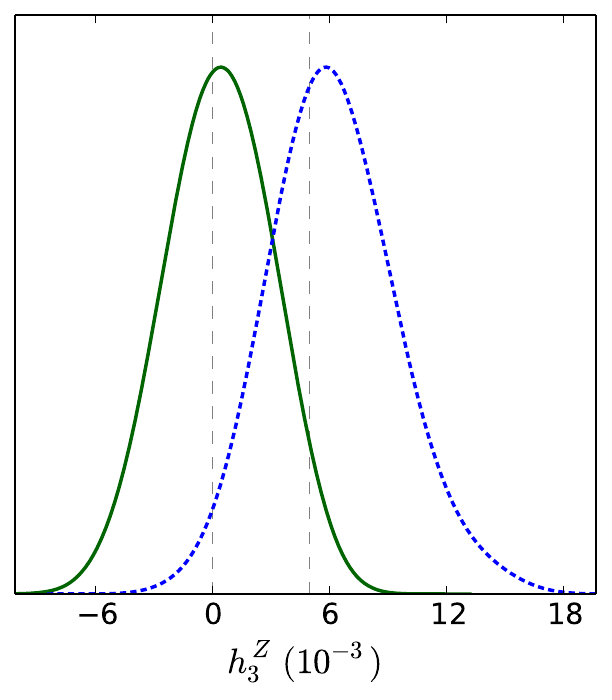}
\caption{\label{fig:one_parameter_mcmc_parameter_za} 
Posterior one-dimensional marginalized distributions for parameters of the
process $e^+e^-\rightarrow Z\gamma$ for {\tt SM} ({\em green/solid}) and
{\tt aTGC} ({\em blue/dashed}) points. Vertical lines denote the values of the
benchmark points }
\end{figure*}
\begin{figure*}
\centering
\includegraphics[width=4.9cm]{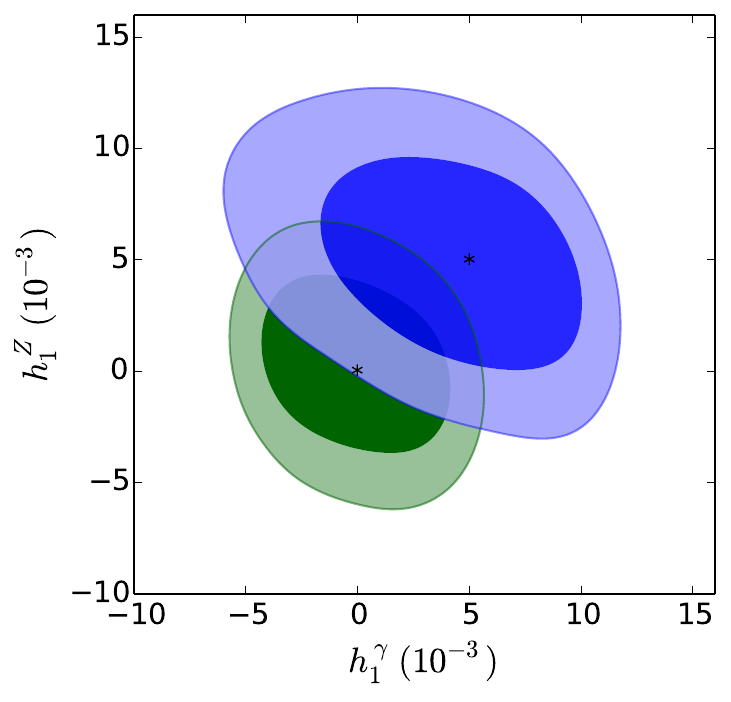}
\includegraphics[width=4.9cm]{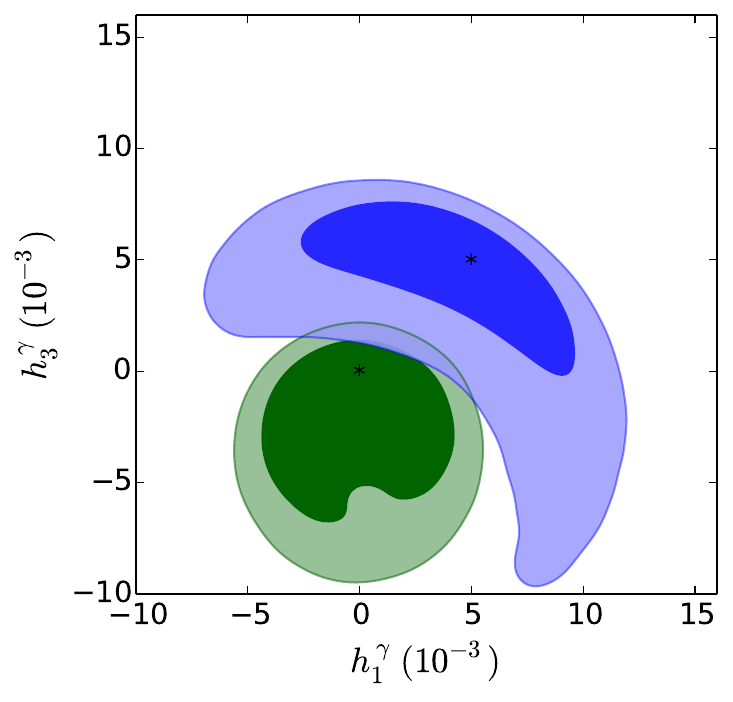}
\includegraphics[width=4.9cm]{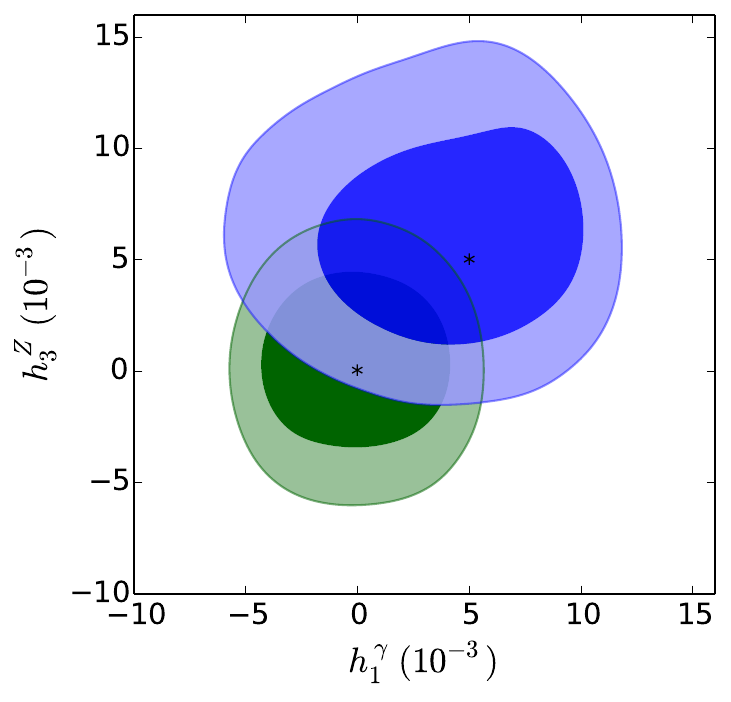}
\includegraphics[width=4.9cm]{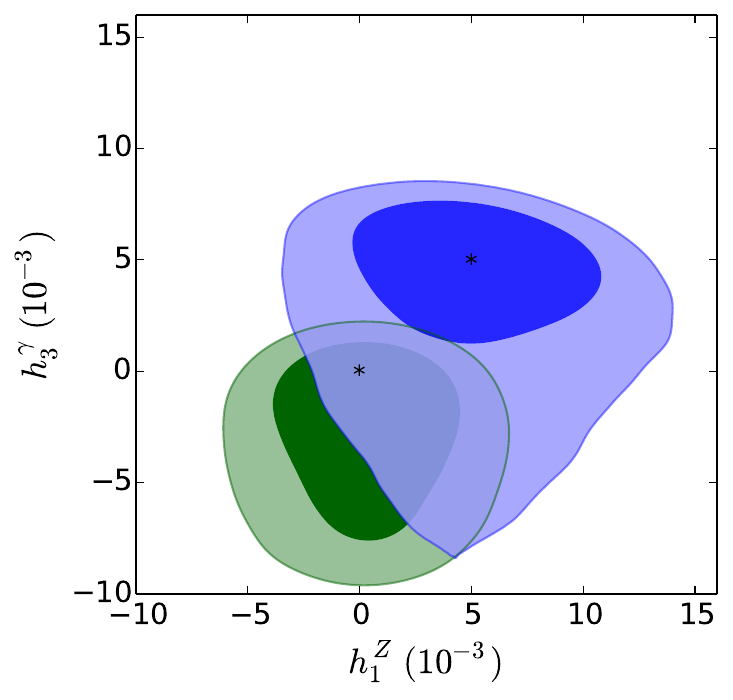}
\includegraphics[width=4.9cm]{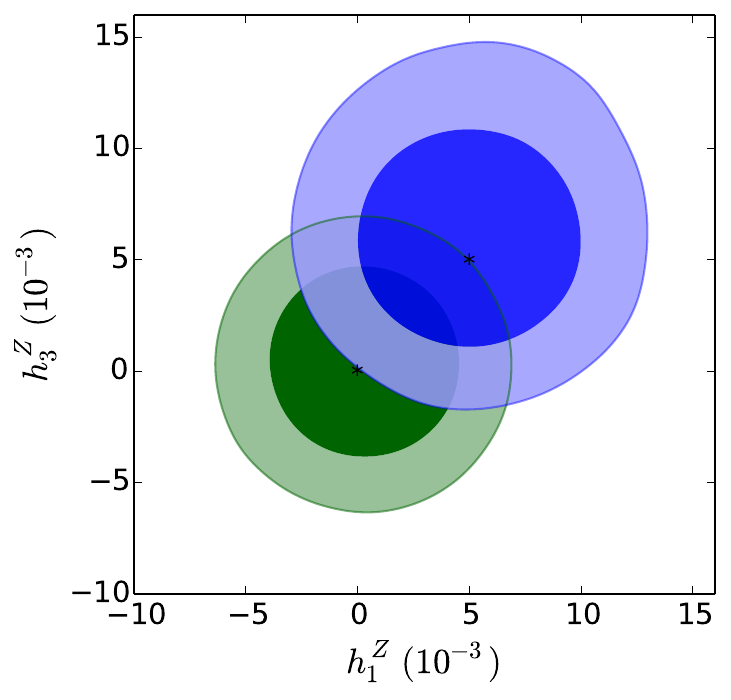}
\includegraphics[width=4.9cm]{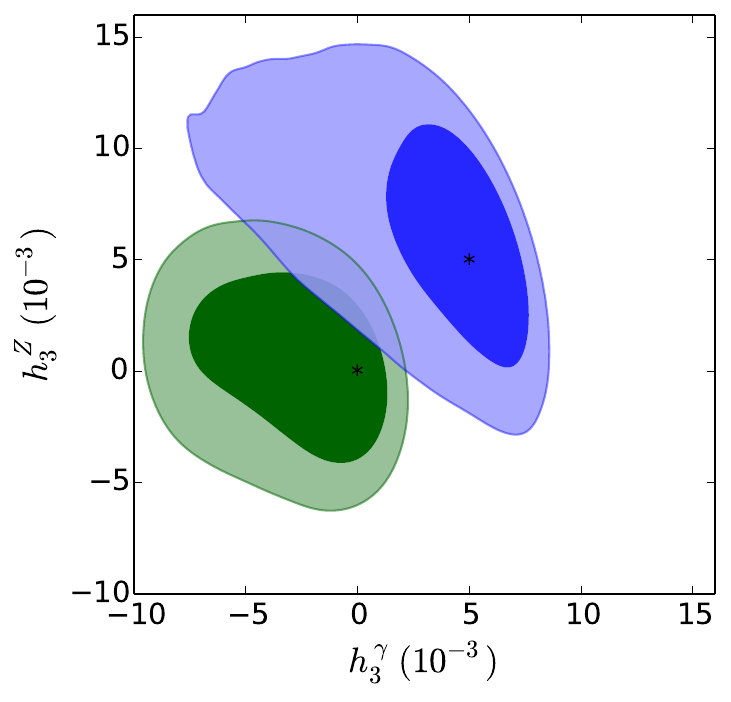}
\caption{\label{fig:correlations_parameters_in_mcmc_za} 
Two-dimensional contours for all pairs of the parameters in the  process 
$e^+e^-\rightarrow Z\gamma$. The Upper transparent layers ({\em blue}) are for
{\tt aTGC} and the lower layers ({\em green}) for the {\tt SM} showing the 68~\% BC 
({\em dark shades}) and  95~\% BC ({\em light shades}) contours}
\end{figure*}
For the parameter extraction we look at their one-dimensional marginalized 
posterior distribution function shown in 
Fig.~\ref{fig:one_parameter_mcmc_parameter_za} for the 
two benchmark points. The best-fit points along with 68~\% and 95~\% BCI are listed
in Table~\ref{tab:marginalised_parameters_za_sm}. The best-fit points are very
close to the true values of the parameters and so are the means of the BCI for
all parameters except $h_3^\gamma$. For it there is a downward movement in the
value owing to the multi-valuedness of the cross section. Also, we note that
the 95~\% BCI for the two benchmark points largely overlap making them seemingly
un-distinguishable at the level of one-dimensional BCIs. To highlight the 
difference between two benchmark points, we look at two-dimensional BC contours
as shown in Fig.~\ref{fig:correlations_parameters_in_mcmc_za}. The 68~\% BC
contours (dark shades) can be roughly compared with the contours of 
Fig.~\ref{fig:Two_parameter_sensitivity_za}. The difference is that 
Fig.~\ref{fig:correlations_parameters_in_mcmc_za} has all four parameters 
varying and all six observables are used simultaneously. The 95~\% BC contours
for the two benchmark points overlap despite the fact that the cross section can
distinguish them very clearly. In full four-dimensional parameter space
the two contours do not overlap and in the next section we try to establish this.

\subsection{Separability of benchmark points }
\label{ssec:separation}
\begin{figure*}
\centering
\includegraphics[width=7.43cm]{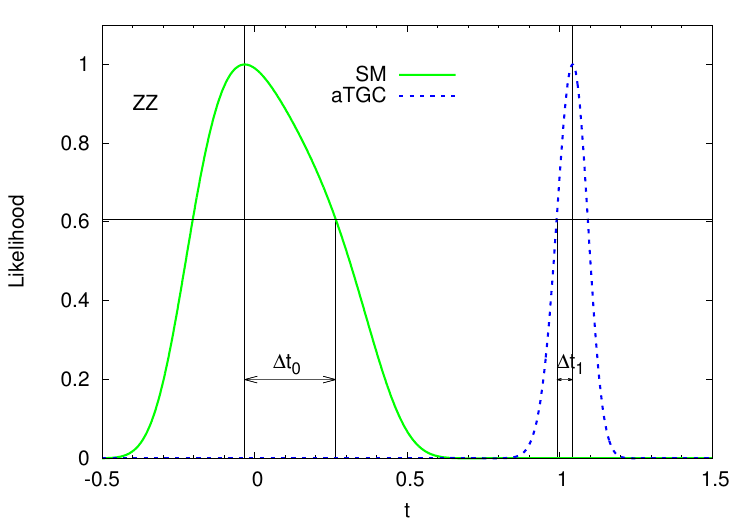}
\includegraphics[width=7.43cm]{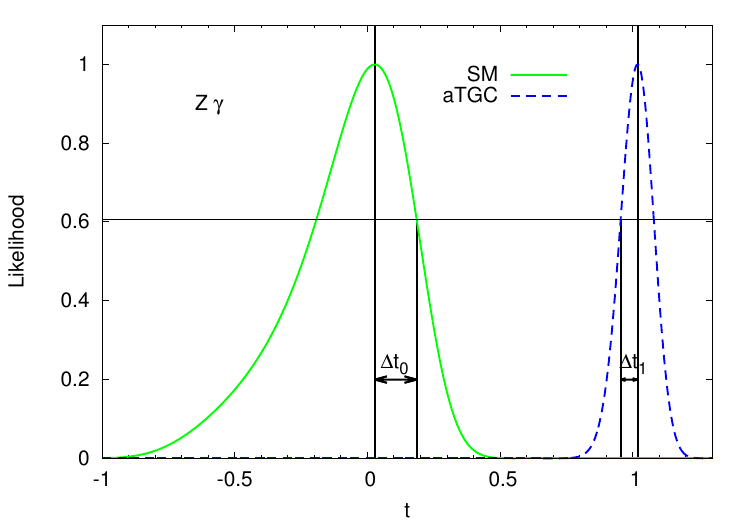}
\caption{\label{fig:likelyhood_ratio} Likelihood ratio for the separability of 
benchmark points for $ZZ$ (left) and  $Z\gamma$ (right) final state: {\tt SM} 
{\em pseudo data} are in solid ({\em green}) and {\tt aTGC} {\em pseudo data} are in 
dotted ({\em blue}) lines}
\end{figure*}
To depict the separability of the two benchmark points pictorially, we vary
all four parameters for a chosen process as a linear function of one parameter,
$t$, as
\begin{equation}
\vec{f}(t) = (1-t) \ \vec{f}_{\tt SM} + t \ \vec{f}_{\tt aTGC},
\end{equation}
such that $\vec{f}(0) = \vec{f}_{\tt SM}$ is the coupling for the
{\tt SM} benchmark point and $\vec{f}(1) = \vec{f}_{\tt aTGC}$ is the
coupling for the {\tt aTGC} point. In Fig.~\ref{fig:likelyhood_ratio} we show
the normalized likelihood for the point $\vec{f}(t)$ assuming the {\tt SM}
{\em pseudo data}, ${\cal L}(\vec{f}(t)|\mbox{\tt SM})$, in  solid/green line
and assuming the {\tt aTGC} {\em pseudo data},
${\cal L}(\vec{f}(t)|\mbox{\tt aTGC})$, in dashed/blue line.
The left panel is for the  $ZZ$ production process and  the right panel
is for the $Z\gamma$ process. The horizontal lines correspond to the normalized
likelihood being $e^{-\frac{1}{2}}$, while the full vertical lines correspond to
the maximum value, which is normalized to $1$. It is clearly visible that the
two benchmark points are quite well separated in terms of the likelihood
ratios. We have ${\cal L}(\vec{f}_{\tt aTGC}|\mbox{\tt SM})\sim 8.8 \times
10^{-19}$
for the $ZZ$ process, and it means that the relative likelihood for the {\tt SM}
{\em pseudo data} being generated by the {\tt aTGC} parameter value is
$8.8 \times 10^{-19}$, i.e. negligibly small. Comparing the likelihood ratio to
$e^{-n^2/2}$ we can say that the data is $n\sigma$ away from the model point.
In this case, {\tt SM} {\em pseudo data} is $9.1\sigma$ away from the
{\tt aTGC} point for the $ZZ$ process. Similarly we have
${\cal L}(\vec{f}_{\tt SM}|\mbox{\tt aTGC})\sim 1.7 \times 10^{-17}$, i.e. the
{\tt aTGC} {\em pseudo data} is $8.8\sigma$ away from the {\tt SM} point for the 
$ZZ$ process.
For the $Z\gamma$ process we have
 ${\cal L}(\vec{f}_{\tt aTGC}|\mbox{\tt SM})\sim 1.7 \times 10^{-24}
(10.5\sigma)$ and
 ${\cal L}(\vec{f}_{\tt SM}|\mbox{\tt aTGC})\sim 1.8 \times 10^{-25}
(10.7\sigma)$.
In all cases the two benchmark points are well separable as clearly seen in 
Fig.~\ref{fig:likelyhood_ratio}.

\section{Conclusions}
\label{sec:conclusion}
There are angular asymmetries in collider that can be
constructed to probe all eight polarization parameters of a massive spin-$1$
particle. Three of them,
$A_y$, $A_{xy}$, and $A_{yz}$, are CP-odd and can be used to measure CP-violation 
in the production process. On the other hand $A_z$, $A_{xz}$, and $A_{yz}$ are 
P-odd observables, while $A_x$, $A_{x^2-y^2}$, and $A_{zz}$ are CP- and P-even.
The anomalous trilinear gauge coupling in the neutral sector,
Eq.~(\ref{aTGC_Lagrangian}), is studied using these asymmetries along with the cross section. The one
and two parameter sensitivity of these asymmetries, together with cross section,
are explored and the one-parameter limit using one observable is listed in 
Table~\ref{tab:aTGC_constrain_form_1sigma_sensitivity} for an unpolarized
$e^+e^-$ collider. For finding the best and simultaneous limit on anomalous
couplings, we have performed a likelihood mapping using the MCMC method and the
obtained limits are listed in Tables~\ref{tab:marginalised_parameters_zz} and
\ref{tab:marginalised_parameters_za_sm} for $ZZ$ and $Z\gamma$ processes,
respectively. 
For the $ZZ$ process, the ILC ($\sqrt{\hat{s}}=500$ GeV, 
${\cal L}=100$ fb$^{-1}$) limits are tighter than the available LHC ($\sqrt{\hat{s}}=7$
TeV, $L=5$ fb$^{-1}$)
limits~\cite{Chatrchyan:2012sga}, while the ILC limits on the $Z\gamma$ anomalous
couplings are slightly weaker than the available LHC ($\sqrt{\hat{s}}=8$
TeV, $L=19.6$ fb$^{-1}$) limits~\cite{Khachatryan:2016yro}. The LHC
probes the interactions at large energies and transverse momentum, where the
sensitivity to the anomalous couplings is enhanced.  We perform our
analysis at $\sqrt{\hat{s}}=500$ GeV, leading to a weaker, though comparable,
limits on the $h_i^V$ in the $Z\gamma$ process.

With polarized initial beams, the P-odd observables, $A_z$, $A_{xz}$, and
$A_{yz}$, will be non-vanishing for both  process with
appropriate kinematical cuts. This gives us three more
observables to add in the likelihood analysis, which can lead to better limits.
At LHC, one does not have the possibility of initial beam polarization, 
however, $W^+W^-$ and $ZW^\pm$ processes effectively have initial beam
polarization due to chiral couplings of $W^\pm$. The study of $W$ and $Z$ processes
at LHC and polarized $e^+e^-$ colliders are underway and will be presented
elsewhere. 


\vspace{0.5cm}
\noindent \textbf{Acknowledgements:} R.~R. thanks Department of Science 
and Technology, Government of India for support through DST-INSPIRE Fellowship 
for doctoral program, INSPIRE CODE IF140075, 2014.  
\appendix
\section{Helicity amplitudes}\label{appendix:helicity_amplitude}
Vertices in SM are taken as
\begin{eqnarray}
e^- e^+ Z^\mu \Rightarrow && -i\frac{g_z}{2} \gamma^\mu \left(C_L P_L + C_R P_R\right),\nonumber \\
e^- e^+ \gamma^\mu \Rightarrow && i g_e \gamma^\mu ,
\end{eqnarray}
where $C_L=-1+2\sin^2\theta_W$, $C_R=2\sin^2\theta_W$, 
 with $\sin^2\theta_W=1-\left(\frac{M_W}{M_Z}\right)^2$. 
 Here $\theta_W$ is the Weinberg mixing angle.
The couplings $g_e$, $g_z$ are given by
\begin{equation}
g_e=\sqrt{4\pi\alpha_{em}} \hspace{1cm}\text{and}\hspace{1cm} 
g_z=\frac{g_e}{\cos\theta_W \sin\theta_W}.
\end{equation}
$P_L=\frac{1-\gamma_5}{2}$, $P_R=\frac{1+\gamma_5}{2}$ are the left and right chiral operators.
Here $\sin\theta$ and $\cos\theta$ are written as $s_\theta$ and $c_\theta$ respectively.
\subsection{For the process $e^+e^-\to ZZ$}

We define  $\beta$  as 
\begin{equation}
\beta=\sqrt{1-\frac{4M_Z^2}{\hat{s}}},
\end{equation}
$\sqrt{\hat{s}}$ being the centre-of-mass energy of the colliding beams.
We also define
$f^Z=f_4^Z+if_5^Z\beta$, 
$f^\gamma=f_4^\gamma+if_5^\gamma\beta$.

\begin{eqnarray}
\mathcal{M}_{\lambda_{e^-},\lambda_{e^+},\lambda_{z(1)},\lambda_{z(2)}}= \mathcal{M}_{SM}+\mathcal{M}_{aTGC} \nonumber
\end{eqnarray}
\begin{eqnarray}
\mathcal{M}_{+,-,+,+} & =&-\frac{ig_z^2C_R^2 \left(1-\beta ^2\right) c_{\theta } s_{\theta }  }{1+2\beta ^2\left(1-2 c_{\theta }^2\right)+\beta ^4}\nonumber\\
\mathcal{M}_{+,-,+,0} & =&
- \frac{i g_z^2C_R^2 \sqrt{1-\beta ^2 }  \left(1-2 c_{\theta }+\beta ^2\right)(1+c_{\theta })  }{\sqrt{2}\left(  1+2\beta ^2\left(1-2 c_{\theta }^2\right)+\beta ^4\right)} \nonumber\\
&&-\frac{\sqrt{2} \left(1+c_{\theta }\right) g_e \beta  \left(2 g_e  f^\gamma-C_R g_z f^Z\right)}{\left(1-\beta ^2\right)^{3/2}}\nonumber\\
\mathcal{M}_{+,-,+,-} & =&
\frac{ig_z^2C_R^2 \left(1+\beta ^2\right) (1+c_{\theta })  s_{\theta }  }{1+2\beta ^2\left(1-2 c_{\theta }^2\right)+\beta ^4} \nonumber\\
\mathcal{M}_{+,-,0,+} & =&
 \frac{i g_z^2C_R^2 \sqrt{1-\beta ^2 }  \left(1+2 c_{\theta }+\beta ^2\right)(1-c_{\theta })  }{\sqrt{2}\left(  1+2\beta ^2\left(1-2 c_{\theta }^2\right)+\beta ^4\right)}\nonumber\\
&&+\frac{\sqrt{2} \left(1-c_{\theta }\right) g_e \beta  \left(2 g_e  f^\gamma-C_R g_z f^Z\right)}{\left(1-\beta ^2\right)^{3/2}}\nonumber\\
\mathcal{M}_{+,-,0,0} & =&
-\frac{2ig_z^2C_R^2 \left(1-\beta ^2\right) c_{\theta } s_{\theta }  }{1+2\beta ^2\left(1-2 c_{\theta }^2\right)+\beta ^4}\nonumber
\end{eqnarray}
\begin{eqnarray}
\mathcal{M}_{+,-,0,-} & =&
- \frac{i g_z^2C_R^2 \sqrt{1-\beta ^2 }  \left(1-2 c_{\theta }+\beta ^2\right)(1+c_{\theta })  }{\sqrt{2}\left(  1+2\beta ^2\left(1-2 c_{\theta }^2\right)+\beta ^4\right)}\nonumber\\
&&+\frac{\sqrt{2} \left(1+c_{\theta }\right) g_e \beta  \left(2 g_e  {f^\gamma}^\star-C_R g_z {f^Z}^\star \right)}{\left(1-\beta ^2\right)^{3/2}}\nonumber\\
\mathcal{M}_{+,-,-,+} & =&
-\frac{ig_z^2C_R^2 \left(1+\beta ^2\right) (1-c_{\theta })  s_{\theta }  }{1+2\beta ^2\left(1-2 c_{\theta }^2\right)+\beta ^4}\nonumber\\
\mathcal{M}_{+,-,-,0} & =&
 \frac{i g_z^2C_R^2 \sqrt{1-\beta ^2 }  \left(1+2 c_{\theta }+\beta ^2\right)(1-c_{\theta })  }{\sqrt{2}\left(  1+2\beta ^2\left(1-2 c_{\theta }^2\right)+\beta ^4\right)}\nonumber\\
&&-\frac{\sqrt{2} \left(1-c_{\theta }\right) g_e \beta  \left(2 g_e  {f^\gamma}^\star-C_R g_z {f^Z}^\star \right)}{\left(1-\beta ^2\right)^{3/2}}\nonumber\\
\mathcal{M}_{+,-,-,-} & =&
-\frac{ig_z^2C_R^2 \left(1-\beta ^2\right) c_{\theta } s_{\theta }  }{1+2\beta ^2\left(1-2 c_{\theta }^2\right)+\beta ^4}\nonumber
\end{eqnarray}
\begin{eqnarray}
\mathcal{M}_{-,+,+,+} & =&
-\frac{ig_z^2C_L^2 \left(1-\beta ^2\right) c_{\theta } s_{\theta }  }{1+2\beta ^2\left(1-2 c_{\theta }^2\right)+\beta ^4}\nonumber\\
\mathcal{M}_{-,+,+,0} & =&
-\frac{i g_z^2C_L^2 \sqrt{1-\beta ^2 }  \left(1+2 c_{\theta }+\beta ^2\right)(1-c_{\theta })  }{\sqrt{2}\left(  1+2\beta ^2\left(1-2 c_{\theta }^2\right)+\beta ^4\right)} \nonumber\\
&&+\frac{\sqrt{2} \left(1-c_{\theta }\right) g_e \beta  \left(2 g_e  f^\gamma- C_L g_z f^Z \right)}{\left(1-\beta ^2\right)^{3/2}}  \nonumber
\end{eqnarray}
\begin{eqnarray}
\mathcal{M}_{-,+,+,-} & =&
\frac{ig_z^2C_L^2 \left(1+\beta ^2\right) (1-c_{\theta })  s_{\theta }  }{1+2\beta ^2\left(1-2 c_{\theta }^2\right)+\beta ^4} \nonumber\\
\mathcal{M}_{-,+,0,+} & =&
\frac{i g_z^2 C_L^2 \sqrt{1-\beta ^2 }  \left(1-2 c_{\theta }+\beta ^2\right)(1+c_{\theta })  }{\sqrt{2}\left(  1+2\beta ^2\left(1-2 c_{\theta }^2\right)+\beta ^4\right)}\nonumber\\
&&-\frac{\sqrt{2} \left(1+c_{\theta }\right) g_e \beta  \left(2 g_e  f^\gamma-C_L g_z f^Z \right)}{\left(1-\beta ^2\right)^{3/2}}\nonumber\\
\mathcal{M}_{-,+,0,0} & =&
-\frac{2ig_z^2C_L^2 \left(1-\beta ^2\right) c_{\theta } s_{\theta }  }{1+2\beta ^2\left(1-2 c_{\theta }^2\right)+\beta ^4}\nonumber
\end{eqnarray}
\begin{eqnarray}
\mathcal{M}_{-,+,0,-} & =&
- \frac{i g_z^2C_L^2 \sqrt{1-\beta ^2 }  \left(1+2 c_{\theta }+\beta ^2\right)(1-c_{\theta })  }{\sqrt{2}\left(  1+2\beta ^2\left(1-2 c_{\theta }^2\right)+\beta ^4\right)}\nonumber\\
&&-\frac{\sqrt{2} \left(1-c_{\theta }\right) g_e \beta  \left(2 g_e  {f^\gamma}^\star - C_L g_z {f^Z}^\star \right)}{\left(1-\beta ^2\right)^{3/2}}\nonumber
\end{eqnarray}
\begin{eqnarray}
\mathcal{M}_{-,+,-,+} & =&
\frac{ig_z^2C_L^2 \left(1+\beta ^2\right) (1+c_{\theta })  s_{\theta }  }{1+2\beta ^2\left(1-2 c_{\theta }^2\right)+\beta ^4} \nonumber
\end{eqnarray}
\begin{eqnarray}
\mathcal{M}_{-,+,-,0} & =&
 \frac{i g_z^2C_L^2 \sqrt{1-\beta ^2 }  \left(1-2 c_{\theta }+\beta ^2\right)(1+c_{\theta })  }{\sqrt{2}\left(  1+2\beta ^2\left(1-2 c_{\theta }^2\right)+\beta ^4\right)}\nonumber\\
&&+\frac{\sqrt{2} \left(1+c_{\theta }\right) g_e \beta  \left(2 g_e  {f^\gamma}^\star -C_L g_z {f^Z}^\star \right)}{\left(1-\beta ^2\right)^{3/2}}\nonumber
\end{eqnarray}
\begin{eqnarray}
\mathcal{M}_{-,+,-,-} & =&
-\frac{i g_z^2C_L^2 \left(1-\beta ^2\right) c_{\theta } s_{\theta }  }{1+2\beta ^2\left(1-2 c_{\theta }^2\right)+\beta ^4}
\end{eqnarray}

\subsection{For the process $e^+e^-\to Z\gamma$}
In this process $\beta$ is given by
\begin{equation}
\beta=1-\frac{M_Z^2}{\hat{s}}.
\end{equation}

We define $h^\gamma=h_1^\gamma + i h_3^\gamma$ and $h^Z=h_1^Z+ih_3^Z$.

\begin{eqnarray}
\mathcal{M}_{ \lambda_{e^-},\lambda_{e^+},\lambda_{z},\lambda_{\gamma}} &= & \mathcal{M}_{SM}+\mathcal{M}_{aTGC} \nonumber
 \end{eqnarray}
 \begin{eqnarray}
 \mathcal{M}_{+,-,+,+}&= &
 \frac{-i  g_e g_z C_Rs_{\theta } (1-\beta )}{\beta (1-c_{\theta }) } \nonumber\\
&&-\frac{g_e \left( 2g_e h^\gamma - C_Rg_z h^Z \right) s_{\theta } \beta }{4 (1-\beta
)}\nonumber
 \end{eqnarray}
 \begin{eqnarray}
  \mathcal{M}_{+,-,+,-}&=&
 \frac{i g_e g_zC_R  s_{\theta }}{\beta(1 -  c_{\theta })}\nonumber
  \end{eqnarray}
 \begin{eqnarray}
  \mathcal{M}_{+,-,0,+}&=&
\frac{i \sqrt{2}g_e g_zC_R\sqrt{1- \beta }  }{\beta }\nonumber\\
&& +\frac{\left(1-c_{\theta }\right)g_e\left(2g_e h^\gamma - C_Rg_z h^Z \right)
\beta }{4 \sqrt{2} (1-\beta )^{3/2}}\nonumber
 \end{eqnarray}
 \begin{eqnarray}
   \mathcal{M}_{+,-,0,-}&=&
 \frac{-i \sqrt{2}g_e g_zC_R\sqrt{1- \beta }  }{\beta }\nonumber\\
&&+\frac{\left(1+c_{\theta }\right)g_e\left(  2g_e {h^\gamma}^\star -C_Rg_z {h^Z}^\star \right)
\beta }{4 \sqrt{2} (1-\beta )^{3/2}} \nonumber
 \end{eqnarray}
 \begin{eqnarray}
   \mathcal{M}_{+,-,-,+}&=&
\frac{-i g_e g_z C_R s_{\theta }}{\beta(1 +  c_{\theta })}\nonumber\\
  \mathcal{M}_{+,-,-,-}&=&
\frac{i g_e g_z (1-\beta )  s_{\theta }}{\beta  (1+c_{\theta })}\nonumber\\
&&-\frac{g_e \left(  2g_e {h^\gamma}^\star -C_Rg_z {h^Z}^\star \right) s_{\theta } \beta }{4 (1-\beta
)}\nonumber
 \end{eqnarray}
 \begin{eqnarray}
   \mathcal{M}_{-,+,+,+}&=&
 \frac{i  g_e g_z C_Ls_{\theta } (1-\beta )}{ \beta ({1+\text{c}_{\theta} })}\nonumber\\ 
&&-\frac{g_e \left( 2g_e h^\gamma - C_L g_z h^Z \right) s_{\theta } \beta }{4
(1-\beta )}\nonumber
 \end{eqnarray}
 \begin{eqnarray}
 \mathcal{M}_{-,+,+,-}&=&
 \frac{-i g_e g_zC_Ls_{\theta }}{\beta(1 +c_{\theta })}\nonumber\\
  \mathcal{M}_{-,+,0,+}&=&
 \frac{i \sqrt{2}g_e g_zC_L\sqrt{1- \beta }  }{\beta }\nonumber\\
&&-\frac{\left(1+c_{\theta }\right)g_e\left(2g_e h^\gamma - C_L g_z h^Z \right)
\beta }{4 \sqrt{2} (1-\beta )^{3/2}} \nonumber
 \end{eqnarray}
 \begin{eqnarray}
 \mathcal{M}_{-,+,0,-}&=&
 \frac{-i \sqrt{2}g_e g_zC_L\sqrt{1- \beta }  }{\beta }\nonumber\\
&& -\frac{\left(1-c_{\theta }\right)g_e\left(2g_e{h^\gamma}^\star -C_L g_z {h^Z}^\star \right)
\beta }{4 \sqrt{2} (1-\beta )^{3/2}}\nonumber\\
\mathcal{M}_{-,+,-,+}&=&
\frac{i g_e g_z C_L s_{\theta }}{\beta(1 - c_{\theta })}\nonumber
  \end{eqnarray}
 \begin{eqnarray}
 \mathcal{M}_{-,+,-,-}&=&
 \frac{-i g_e g_zC_L (1-\beta )  s_{\theta }}{\beta  (1-c_{\theta })}\nonumber\\
&& -\frac{g_e \left(2g_e{h^\gamma}^\star -C_L g_z {h^Z}^\star \right) s_{\theta } \beta }{4
(1-\beta )}
\end{eqnarray}

\section{Polarization observables}\label{appendix:Expresson_observables}
\subsection{For the process $e^+e^-\to ZZ$}
\begin{equation}
\sigma(e^+e^-\to Z Z)=\frac{1}{2} \dfrac{1}{32\pi\beta \hat{s}} \ 
\tilde{\sigma}_{ZZ}
\end{equation}

\begin{eqnarray}
&&\tilde{\sigma}_{ZZ}=\frac{1}{2} g_Z^4\left(C_L^4+C_R^4\right)\left(\frac{\left(5-2 \beta ^2+\beta ^4\right) \log\left(\frac{1+\beta
}{1-\beta }\right)}{2 \left(\beta +\beta ^3\right)}-1\right)\nonumber\\
&&+ g_Z^3 g_e \left(C_L^3-C_R^3\right) f_5^Z\frac{ \left(3-\beta ^2\right)\left(2 \beta  +\left(1+\beta ^2\right) \log\left(\frac{1+\beta
}{1-\beta }\right)\right)}{2 \beta  \left(1-\beta ^2\right)}\nonumber\\
&&- g_Z^2 g_e^2\left(C_L^2-C_R^2\right) f_5^{\gamma } \frac{ \left(3-\beta ^2\right)\left(2 \beta  -\left(1+ \beta ^2\right) \log\left(\frac{1+\beta
}{1-\beta }\right)\right)}{\beta  \left(1-\beta ^2\right)}\nonumber\\
&&+ 16 g_Z^2 g_e^2\left(C_L^2+C_R^2\right)\frac{\beta ^2 \left((f_4^Z)^2+(f_5^Z)^2 \beta ^2\right)}{3 \left(1-\beta ^2\right)^3}\nonumber\\
&&- 64 g_Z g_e^3 \left(C_L+C_R\right)\frac{ \beta ^2 \left(f_4^{\gamma } f_4^Z+f_5^{\gamma } f_5^Z \beta ^2\right)}{3 \left(1-\beta ^2\right)^3}\nonumber\\
&&+ 128 g_e^4\frac{ \beta ^2 \left((f_4^{\gamma })^2+(f_5^{\gamma })^2 \beta ^2\right)}{3 \left(1-\beta ^2\right)^3}
\end{eqnarray}
\begin{eqnarray}
\tilde{\sigma}_{ZZ}\times P_x =&& g_Z^4 \left(C_L^4-C_R^4\right)\frac{\pi(1-\beta^2)^{3/2}}{4(1+\beta^2)}\nonumber\\
&+& g_z^3 g_e \left(C_L^3+C_R^3\right)\frac{f_5^Z\pi\beta^2(1+2\beta^2)}{4(1-\beta^2)^{3/2}}\nonumber\\
&-& g_z^2 g_e^2 \left(C_L^2+C_R^2\right)\frac{f_5^\gamma\pi\beta^2(1+2\beta^2)}{2(1-\beta^2)^{3/2}}
\end{eqnarray}
\begin{eqnarray}
\tilde{\sigma}_{ZZ}\times P_y = 
&-& g_z^3 g_e \left(C_L^3+C_R^3\right)\frac{f_4^Z\pi\beta(2+\beta^2)}{4(1-\beta^2)^{3/2}}\nonumber\\
&+& g_z^2 g_e^2 \left(C_L^2+C_R^2\right)\frac{f_4^\gamma\pi\beta(2+\beta^2)}{2(1-\beta^2)^{3/2}}
\end{eqnarray}
\begin{eqnarray}
&&\tilde{\sigma}_{ZZ}\times T_{xy}=
g_Z^3g_e\left(C_L^3-C_R^3\right)\frac{\sqrt{3}}{8}f_4^Z f_{xy}(\beta)\nonumber\\
&&-g_Z^2g_e^2 \left(C_L^2-C_R^2\right)\frac{\sqrt{3}}{4}f_4^{\gamma }f_{xy}(\beta)\nonumber\\
&&+g_Z^2 g_e^2 \left(C_L^2+C_R^2\right)\frac{4}{\sqrt{3}}\frac{ f_4^Z f_5^Z\beta ^3}{\left(1-\beta ^2\right)^3}\nonumber\\
&&+g_Z g_e^3\left(C_L+C_R\right)\frac{8}{\sqrt{3}}\frac{ \left(f_4^Z f_5^{\gamma }+f_4^{\gamma } f_5^Z\right)\beta ^3}{\left(1-\beta
^2\right)^3}\nonumber\\
&&+ g_e^4\frac{32f_4^{\gamma } f_5^{\gamma }\beta ^3}{\sqrt{3} \left(1-\beta ^2\right)^3},\nonumber\\
&&f_{xy}(\beta)=\frac{\left(2 \left(\beta +\beta ^3\right)-\left(1-\beta
^2\right)^2 \log\left(\frac{1+\beta }{1-\beta }\right)\right)}{ \beta ^2 \left(1-\beta ^2\right)}
\end{eqnarray}
\begin{eqnarray}
&&\tilde{\sigma}_{ZZ}\times(T_{xx}-T_{yy})=\nonumber\\
&&g_Z^4 \left(C_L^4+C_R^4\right) \sqrt{\frac{3}{2}}\frac{ \left(1-\beta ^2\right) \left(2 \beta -\left(1+\beta ^2\right) \log\left(\frac{1+\beta
}{1-\beta }\right)\right)}{8 \beta ^3}\nonumber\\
&&-\bigg( g_Z^3g_e \left(C_L^3-C_R^3\right) f_5^Z
+ g_Z^2g_e^2\left(C_L^2-C_R^2\right) 2f_5^{\gamma }\bigg) f_{x^2-y^2}(\beta)  \nonumber\\
&&+ g_Z^2g_e^2 \left(C_L^2+C_R^2\right)2 \sqrt{\frac{2}{3}}\frac{\beta ^2 \left((f_4^Z)^2-(f_5^Z)^2 \beta ^2\right)}{\left(1-\beta
^2\right)^3}\nonumber\\
&&- g_Z g_e^3\left(C_L+C_R\right)8 \sqrt{\frac{2}{3}}\frac{\beta ^2 \left(f_4^{\gamma } f_4^Z-f_5^{\gamma } f_5^Z \beta ^2\right)}{\left(1-\beta
^2\right)^3}\nonumber\\
&&+ g_e^416 \sqrt{\frac{2}{3}}\frac{\beta ^2 \left((f_4^{\gamma })^2-(f_5^{\gamma })^2 \beta ^2\right)}{\left(1-\beta ^2\right)^3},\nonumber\\
&&f_{x^2-y^2}(\beta)=\sqrt{\frac{3}{2}}\frac{\left(2 \left(\beta +\beta ^3\right)-\left(1-\beta ^2\right)^2
\log\left(\frac{1+\beta
}{1-\beta }\right)\right)}{4 \beta  \left(1-\beta ^2\right)}\nonumber\\
\end{eqnarray}
\begin{eqnarray}
&&\tilde{\sigma}_{ZZ}\times T_{ZZ}=
g_Z^4 \left(C_L^4+C_R^4\right)\frac{f_{zz}(\beta) }{8 \sqrt{6} \beta ^3 \left(1+\beta ^2\right)}\nonumber\\
&&+ g_Z^3g_e \left(C_L^3-C_R^3\right) f_5^Z\frac{f_{zz}(\beta) }{4 \sqrt{6} \beta  \left(1-\beta ^2\right)}\nonumber\\
&-&g_Z^2g_e^2 \left(C_L^2-C_R^2\right) f_5^{\gamma } \frac{f_{zz}(\beta) }{2 \sqrt{6} \beta  \left(1-\beta ^2\right)}\nonumber\\
&&-g_Z^2 g_e^2  \left(C_L^2+C_R^2\right) \frac{4}{3} \sqrt{\frac{2}{3}}\frac{\beta ^2 \left((f_4^Z)^2+(f_5^Z)^2 \beta ^2\right)}{
\left(1-\beta ^2\right)^3}\nonumber\\
&&+g_Zg_e^3\left(C_L+C_R\right)\frac{16 }{3}\sqrt{\frac{2}{3}}\frac{\beta ^2 \left(f_4^{\gamma } f_4^Z+f_5^{\gamma }
f_5^Z \beta ^2\right)}{ \left(1-\beta ^2\right)^3}\nonumber\\
&&-g_e^4\frac{32}{3} \sqrt{\frac{2}{3}}\frac{\beta ^2 \left((f_4^{\gamma })^2+(f_5^{\gamma })^2 \beta ^2\right)}{ \left(1-\beta
^2\right)^3},\nonumber\\
&&f_{zz}(\beta)=\bigg(\left(3+\beta ^2+5 \beta ^4-\beta ^6\right) \log\left(\frac{1+\beta }{1-\beta}\right)\bigg.\nonumber\\
&&\hspace{1.5cm}\bigg. -2 \beta  \left(3-\beta ^2\right) \left(1+\beta ^2\right)\bigg)
\end{eqnarray}
\subsection{For the process $e^+e^-\to Z\gamma$}
\begin{equation}
\sigma(e^+e^-\to Z \gamma)=\dfrac{1}{32\pi\beta \hat{s}} \ 
\tilde{\sigma}_{Z\gamma} ,\ \ \ c_\theta\in[-c_{\theta_0},c_{\theta_0}]
\end{equation}

\begin{eqnarray}
&&\tilde{\sigma}_{Z\gamma}=
 - g_e^2 g_Z^2\left(C_L^2+C_R^2\right)\frac{1}{\beta ^2}\bigg(c_{\theta _0} \beta ^2\bigg.\nonumber\\
 &&-\bigg. 2 \left(2-2 \beta +\beta ^2\right) \tanh ^{-1}(c_{\theta_0})\bigg)\nonumber \\
&&+ g_e^2 g_Z^2\left(C_L^2-C_R^2\right)\frac{c_{\theta _0}h_3^Z (2-\beta )}{2 (1-\beta )} \nonumber\\
&&- g_e^3 g_Z \left(C_L-C_R\right)\frac{c_{\theta _0}h_3^{\gamma } (2-\beta )}{(1-\beta) } \nonumber\\
&&+ g_e^2 g_Z^2\left(C_L^2+C_R^2\right)c_{\theta _0}\left((h_1^Z)^2+(h_3^Z)^2\right) h_{\sigma}(\beta) \nonumber\\
&-& g_e^3 g_Z\left(C_L+C_R\right) c_{\theta _0}\left(h_1^{\gamma } h_1^Z+h_3^{\gamma } h_3^Z\right) 4 h_{\sigma}(\beta)\nonumber\\
&&+ g_e^4 c_{\theta _0}\left((h_1^{\gamma })^2+(h_3^{\gamma })^2\right) 8h_{\sigma}(\beta),\nonumber \\
&& h_{\sigma}(\beta)=\frac{\beta ^2 \left(9-6 \beta -c_{\theta
_0}^2 (1-2 \beta )\right)}{96 (1-\beta )^3}
\end{eqnarray}
\begin{eqnarray}
&&\tilde{\sigma}_{Z\gamma}\times P_x =\nonumber\\
&&  g_e^2 g_Z^2\left(C_L^2-C_R^2\right)2\frac{\sqrt{1-\beta }}{\beta ^2} (2-\beta ) \sinh ^{-1}(c_{\theta_0})\nonumber\\
&&+ g_e^2 g_Z^2 \left(C_L^2+C_R^2\right)h_3^Z  h_x^1(\beta)
- g_e^3 g_Z\left(C_L+C_R\right)h_3^{\gamma }  2 h_x^1(\beta)\nonumber\\
&&+ g_e^2 g_Z^2 \left(C_L^2-C_R^2\right) \left((h_1^Z)^2+(h_3^Z)^2\right)h_x^2(\beta) \nonumber\\
&&- g_e^3 g_Z\left(C_L-C_R\right)\left(h_1^{\gamma } h_1^Z+h_3^{\gamma } h_3^Z\right) 4h_x^2(\beta),\nonumber\\
&&h_x^1(\beta)=\frac{ \left(c_{\theta _0} s_{\theta _0} (2-3 \beta )-3 (2-\beta ) \sinh ^{-1}(c_{\theta_0})\right)}{8 (1-\beta )^{3/2}},\nonumber\\
&&h_x^2(\beta)=\frac{ \beta ^2 \left(c_{\theta _0} s_{\theta
_0}+\sinh ^{-1}(c_{\theta_0})\right)}{32 (1-\beta )^{5/2}}
\end{eqnarray}
\begin{eqnarray}
&&\tilde{\sigma}_{Z\gamma}\times P_y=\nonumber\\
&&g_e^2 g_Z^2\left(C_L^2+C_R^2\right) \frac{h_1^Z\bigg(c_{\theta _0}s_{\theta _0}\beta-(4-\beta)\sin^{-1}(c_{\theta _0})\bigg)}{8(1-\beta)^{3/2}}\nonumber\\
&&- g_e^3 g_Z\left(C_L+C_R\right) \frac{h_1^\gamma\bigg(c_{\theta _0}s_{\theta _0}\beta-(4-\beta)\sin^{-1}(c_{\theta _0})\bigg)}{4(1-\beta)^{3/2}}
\end{eqnarray}
\begin{eqnarray}
\tilde{\sigma}_{Z\gamma}\times T_{xy}&=&
g_e^3 g_Z \left(C_L-C_R\right)\frac{\sqrt{3}}{2}\frac{h_1^\gamma c_{\theta _0}}{2(1-\beta)}\nonumber\\
&-& g_e^2 g_Z^2 \left(C_L^2-C_R^2\right)\frac{\sqrt{3}}{4}\frac{h_1^Z c_{\theta _0}}{4(1-\beta)}
\end{eqnarray}
\begin{eqnarray}
\tilde{\sigma}_{Z\gamma}\times(T_{xx}-T_{yy})=&& 
g_e^2 g_Z^2\left(C_L^2+C_R^2\right)\frac{\sqrt{6} c_{\theta _0}(1-\beta)}{\beta^2}\nonumber\\
&-& g_e^3 g_Z \left(C_L-C_R\right)\sqrt{\frac{3}{2}} \frac{h_3^\gamma c_{\theta _0}}{(1-\beta)} 
\end{eqnarray}

\begin{eqnarray}
&&\tilde{\sigma}_{Z\gamma}\times T_{zz}=
- g_e^2 g_Z^2\left(C_L^2+C_R^2\right)\frac{1}{\sqrt{6} \beta ^2}\bigg(c_{\theta _0} \left(6-6 \beta +\beta ^2\right)\bigg.\nonumber\\
&&\bigg.-2 \left(2-2 \beta
+ \beta ^2\right) \tanh ^{-1}(c_{\theta_0})\bigg)\nonumber\\
&&- g_e^2 g_Z^2\left(C_L^2-C_R^2\right)\frac{c_{\theta _0}h_3^Z (1+\beta )}{2 \sqrt{6} (1-\beta )}\nonumber\\
&&+ g_e^3 g_Z\left(C_L-C_R\right)\frac{c_{\theta _0}h_3^{\gamma } (1+\beta )}{\sqrt{6} (1-\beta )}\nonumber\\
&&- g_e^2 g_Z^2\left(C_L^2+C_R^2\right) c_{\theta _0}\left((h_1^Z)^2+(h_3^Z)^2\right) h_{zz}(\beta) \nonumber\\
&&+ g_e^3 g_Z \left(C_L+C_R\right) c_{\theta _0}\left(h_1^{\gamma } h_1^Z+h_3^{\gamma } h_3^Z\right) 4h_{zz}(\beta) \nonumber\\
&&- g_e^4 c_{\theta _0}\left((h_1^{\gamma })^2+(h_3^{\gamma })^2\right) 8h_{zz}(\beta) ,\nonumber\\
&&h_{zz}(\beta)=\frac{ \left(c_{\theta _0}^2 (2-\beta )+3 \beta \right)
\beta ^2}{48 \sqrt{6} (1-\beta )^3}
\end{eqnarray}

\fontsize{9}{10}\selectfont
\bibliography{ATGC_VecPol}
\bibliographystyle{utphys}

\end{document}